\documentclass[10pt,journal,twocolumn]{IEEEtran}

\usepackage[utf8]{inputenc} 
\usepackage[T1]{fontenc}
\usepackage{url}              
\usepackage{cite}             
\usepackage{color}
\usepackage[colorlinks, linkcolor=color1, anchorcolor=blue, citecolor=color1]{hyperref}
\definecolor{color1}{RGB}{128,0,0}

\usepackage{bm,amsfonts}
\usepackage[cmex10]{amsmath}  
\usepackage{mleftright}       
\mleftright                   

\usepackage{graphicx}         
\usepackage{booktabs}         

\usepackage{algorithm,algorithmic}    

\usepackage[caption=false,font=footnotesize]{subfig}

\usepackage{amssymb}

\begin{document}

\title{Active Sensing for Reciprocal MIMO Channels} 


\author{Tao Jiang,~\IEEEmembership{Graduate Student Member,~IEEE,}
    and Wei Yu,~\IEEEmembership{Fellow,~IEEE}
\thanks{ 
Tao Jiang and Wei Yu are with The Edward S.\ Rogers Sr.\ Department of
Electrical and Computer Engineering, University of Toronto, Canada. (e-mails: taoca.jiang@mail.utoronto.ca, weiyu@ece.utoronto.ca). 
This work is supported by the Natural Sciences and Engineering Research Council of Canada via the Canada Research Chairs program. The materials in this work have been presented in part at IEEE International Workshop on Signal Processing Advances in Wireless Communications (SPAWC), Shanghai, China, September 2023 \cite{tao_spawc}. The source code for this paper is available at: \url{https://github.com/taojiang-github/Active-Sensing-for-Reciprocal-MIMO-Channels} }}

\maketitle

\begin{abstract}
This paper addresses the design of transmit precoder and receive combiner 
matrices to support $N_{\rm s}$ independent data streams over a time-division
duplex (TDD) point-to-point massive multiple-input multiple-output (MIMO)
channel with either a fully digital or a hybrid structure. The optimal precoder
and combiner design amounts to finding the top-$N_{\rm s}$ singular vectors of
the channel matrix, but the explicit estimation of the entire high-dimensional 
channel would require significant pilot overhead. Alternatively, prior works
suggest to find the precoding and combining matrices directly by exploiting
channel reciprocity and by using the power iteration method, but its
performance degrades in the low SNR regime.  To tackle this challenging
problem, this paper proposes a learning-based active
sensing framework, where the transmitter and the receiver send pilots
alternately using sensing beamformers that are actively designed as functions
of previously received pilots. This is accomplished by using recurrent neural
networks to summarize information from the historical observations into hidden
state vectors, then using fully connected neural networks to learn the
appropriate sensing beamformers in the next pilot stage and finally the
transmit precoding and receive combiner matrices for data communications.  
Simulations demonstrate that the learning-based method outperforms existing
approaches significantly and maintains superior performance even in the low SNR
regime for both the fully digital and hybrid MIMO scenarios.  
\end{abstract}


\section{Introduction}

Massive multiple-input multiple-output (MIMO) technology is a key enabler
for high spectral efficiency communications in 5G and future wireless systems
\cite{6375940,BJORNSON20193}, particularly in mmWave bands, where a large
number of antennas can be packed into a small volume due to the short
wavelength \cite{6515173,7400949}. Beamforming is essential for reaping the
benefit of massive MIMO. However, the conventional design of the transmit
precoding and receive combiner matrices typically requires accurate
high-dimensional channel state information (CSI). Yet in the pilot stage, the
MIMO channels are often sensed only through low-dimensional observations,
e.g., because of the limited number of beamformed pilots, or because the antenna arrays may be implemented using a hybrid analog
and digital structure with only a small number of radio frequency (RF) chains
\cite{6717211}.  Consequently, the conventional approach of first estimating
the channel then designing the optimal precoding and combiner matrices for the massive MIMO
system would require significant pilot training overhead.

In this paper, we propose an active sensing framework for the design of
transmit precoder and receive combiner in a time-division duplex (TDD)
point-to-point massive MIMO system without explicit channel estimation.
Specifically, we address the problem of supporting $N_{\rm s}$ independent data
streams between two multi-antenna agents, for both the fully digital and the
hybrid antenna structures. The design of the optimal transmit precoder and
receive combiner matrices is essentially that of finding the left and right
singular vectors of the channel matrix corresponding to the top-$N_{\rm s}$
singular values. But instead of first estimating the channel matrix (which
would require many pilots) then performing singular-value decomposition (SVD), 
this paper adopts a strategy in which the two agents utilize an interactive 
pilot phase to sense the channel in a ping-pong fashion based on channel
reciprocity, in which a learning-based approach is used to design the sensing
beamformers sequentially, followed by the eventual precoder and combiner
matrices design based on all the observations.   

In the ping-pong pilot strategy, 
the two agents alternately probe the channel in multiple rounds through 
adaptively designed low-dimensional sensing beamformers at both the transmitter
and the receiver.  The design of the optimal sensing strategy in this context
is however a challenging task, because the two agents need to implicitly cooperate in
designing their transmit and receive beamformers that can provide the most
informative observations about the top-$N_{\rm s}$ singular vectors of the high
dimensional channel, while accounting for their existing knowledge about the
channel. The idea of active sensing has been considered in an early work
\cite{1323251}, which connects the channel sensing problem with the power
iteration method for finding the dominant singular vectors of the channel
matrix.  However, the power iteration method of \cite{1323251} is developed for a noiseless
setting, which limits its applicability to the high signal-to-noise ratio (SNR)
regime. For the hybrid MIMO system, the hybrid analog-digital architecture raises 
additional challenges, because the sensing beamformers and the eventual precoder 
and combiner matrices must all conform to the specific hybrid structure, so 
the conventional power iteration method cannot be directly applied.


The conventional approach to massive MIMO system design involves first estimating the channel and then optimizing the beamformers based on the estimated channel \cite{1193803,9427148}. To reduce the pilot training overhead, the knowledge of the underlying structure of the channel, e.g., sparsity in mmWave channels, can be utilized in the channel estimation step \cite{6847111,8625694,8356247}. However, such an approach relies heavily on the geometry of the antenna arrays and the sparsity assumption on the channel model, which is not necessarily accurate. Moreover, in conventional approaches, the sensing beamformers for pilot transmission are often generated randomly, but sensing in random directions is not necessarily optimal, because typically only a low-dimensional subspace of the channel is of interest for designing the precoder and the combiner. 

Inspired by recent developments in deep learning for reducing pilot training
overhead \cite{9914567}, this paper proposes a learning-based sensing framework
that directly learns the beamformers from received pilots without estimating
the entire channel and in an active fashion over multiple rounds. The main idea
is to utilize a set of carefully designed recurrent neural networks (RNNs) and
fully connected deep neural networks (DNNs) to automatically summarize
information from the historical observations into state vectors, which are used to
design the sensing beamformers in the next stage, so that the sensing beamformers gradually
focus the energy towards the directions of interest, e.g., the directions of the
top-$N_{\rm s}$ singular vectors. We show that the proposed framework can work
well in both fully digital and hybrid beamforming scenarios and is robust to
the noise. 
 
\subsection{Related Works}
Recently, active sensing has shown remarkable performance for various applications including beam alignment \cite{8792366,sohrabi2021active,10051966,10124207}, beam tracking \cite{han2023} and localization \cite{Zhongze2023}. The key idea behind active sensing is to dynamically design the sensing beamformers based on previously received pilots, gradually focusing on the low-dimensional part of the channel. For instance, \cite{sohrabi2021active,Zhongze2023} demonstrate that the active sensing approach can adaptively probe the channel using broad beams initially and then narrow down the searching range for accurate angle-of-arrival estimation or localization. In this paper, we propose an active sensing framework to adaptively design the sensing beamformers to focus the sensing energy towards the top-$N_{\rm s}$ singular vector directions to learn the optimal precoding and beamforming matrices.

This paper is most closely related to the works \cite{10124207,1323251,7947217,4203059} about actively learning the singular vector pairs of the channel matrix. In \cite{10124207}, the focus lies on learning the top singular vector pair of the channel with both the transmitter and receiver having only one RF chain. This paper extends the framework to the scenario with multiple data streams, where multiple singular vectors need to be estimated.  In a fully digital MIMO setup, \cite{1323251,4203059} propose to send pilots back and forth from both sides while the pilot sequences (or equivalently, the sensing beamformers) are designed based on the received pilots to mimic the power iteration methods. The paper \cite{7575663} generalizes the power iteration method to the multiuser MIMO transceiver design, where the power iteration method is utilized to estimate the top left and right singular vectors of the channel, and a subsequent zero-forcing algorithm is used to mitigate the inter-user interference based on the estimated singular vectors. 
However, the algorithms based on the vanilla power method can converge to a highly suboptimal solution in the low SNR regime, which is a typical scenario in the mmWave initial alignment phase. The paper \cite{7947217} proposes some techniques to address this problem, but it introduces extra feedback overhead between both sides.

The power iteration method is also exploited in the design of the hybrid beamformers in \cite{7556971,7439748, 7996580,8644444}. The work \cite{7556971} proposes to apply unit modulus normalization within the power iteration method to design the analog beamformers, but it assumes that the high dimensional signal in the analog domain can be directly observed, which may not be practically feasible. In contrast, the paper \cite{7439748} considers a more realistic scenario where only a low dimensional signal after RF chain processing can be observed. To gather sufficient observations for employing the power iteration method, they introduce a repetition-aided ping-pong pilot protocol, but repetition leads to significant pilot overhead. Moreover, \cite{7439748} needs an additional matrix decomposition step to obtain the separate analog and digital beamformers, which introduces additional computational complexity due to the need to solve a nonconvex optimization problem. The works \cite{7996580,8644444} use the power iteration method for designing digital beamformers, while analog beamformers are adaptively chosen from a predefined codebook. However, these methods are sensitive to noise, and the codebook constraint can limit their performance as well.

\subsection{Main Contributions}
This paper proposes an active sensing framework that achieves superior performance across a wide range of SNRs without introducing extra feedback overhead in both fully digital and hybrid beamforming scenarios. 
Our main contributions are summarized as follows:

\subsubsection{Active Sensing for Fully Digital MIMO Systems}
To support $N_{\rm s}$ independent data streams in a fully digital
point-to-point MIMO channel, this paper proposes an active sensing framework
based on a ping-pong pilot transmission scheme to estimate the top right 
and left singular vectors of the channel, which are used as the precoder at the
transmitter and the combiner at the receiver. Specifically, agent A and
agent B send pilots alternately using sensing beamformers designed by the
proposed active sensing framework. The sensing beamformers are designed based
on previously received pilots using a novel DNN architecture consisting of
parallel gated recurrent units (GRUs), which abstract information from
historical observations, and parallel fully connected deep neural networks,
which design the sensing beamformers and the precoder and combiner matrices.  
Simulations show that the proposed framework works effectively even
for the most challenging Rayleigh fading channel.

\subsubsection{Active Sensing for Hybrid MIMO Systems}
We further extend the proposed active sensing framework to the hybrid
architecture consisting of both the analog and the digital beamforming
components.  To deal with this challenging scenario, we introduce an additional
set of DNNs that leverage all the information from the current hidden state
vectors to design the analog sensing beamformers in the next stage in a
codebook-free manner. In the meanwhile, the digital sensing beamformers are designed
similarly to the fully digital MIMO scenario. The joint design of analog and
digital beamformers eliminates the need for the explicit decomposition of the
overall beamforming matrices, resulting in a more computationally efficient
solution.  Simulation results demonstrate that this proposed approach achieves
remarkable performance gains over the benchmarks.

\subsubsection{Interpretation of Learned Solutions} 
The solutions learned by the proposed active sensing approach are interpretable. 
The proposed approach successfully matches the top $N_{\rm s}$
singular vectors with minimal energy leakage to other singular
vector directions as compared to the benchmarks, both in the fully digital and the hybrid MIMO
scenarios. This interpretability aids in understanding and validating the
effectiveness of the proposed active sensing framework.

\subsection{Paper Organization and Notations}
The remaining part of this paper is organized as follows. Section \ref{sec:digital_MIMO} describes the system model and the problem formulation in the digital MIMO scenario. Section \ref{sec:active_unit} introduces the proposed active sensing framework for the digital MIMO system. Section \ref{sec:simualtion_digital} provides simulation results of the digital MIMO system. 
Section \ref{sec:model_hybrid} describes the system model and problem formulation of the hybrid scenario. Section \ref{sec:method_hybrid} and Section \ref{sec:simulation_hybrid} present the proposed active sensing framework and simulation results for the hybrid beamforming case. Section \ref{sec:interpretation} presents the interpretation of the learned solutions. Section \ref{sec:conclusion} concludes the paper. 

We use lowercase letters (e.g., $a$), lowercase bold-faced letters (e.g., $\bm a$), and uppercase bold-faced letters (e.g., $\bm A$) to denote scalars, vectors, and matrix, respectively. We use $(\cdot)^\top$ and $(\cdot)^{\sf H}$ to denote transpose and Hermitian transpose, respectively. We use $\mathcal{CN}(\cdot,\cdot)$ to denote complex Gaussian distributions. We use $[\bm F]_{ij}$ to denote the element of the $i$-th row and $j$-th column of the matrix $\bm F$.

\section{Active Sensing for Fully Digital MIMO System}\label{sec:digital_MIMO}
\subsection{System Model}
\label{sec:system-model}
The first part of this paper considers a fully digital point-to-point MIMO system, where agent A with $M_{\rm t}$ fully digital antennas communicates with agent B with $M_{\rm r}$ fully digital antennas. The channel matrix from agent A to agent B is denoted as $\bm G\in\mathbb{C}^{M_{\rm r}\times M_{\rm t}}$. We assume the system operates in TDD mode and that the channel reciprocity holds\footnote{\textcolor{black}{In practical systems, calibration is essential to ensure channel reciprocity, compensating for transceiver impairments and temperature variations. However, such calibration is only required on a long-term basis, e.g., hourly.}}, i.e., the uplink channel can be represented as $\bm G^{\sf H}$. \textcolor{black}{We assume a narrowband and frequency-flat block fading channel model, where the channel remains constant within a coherence block but changes independently across different blocks.}
The received signal $\bm y\in\mathbb{C}^{M_{\rm r}}$ at agent B can be written as:
\begin{align}
  \bm y = \bm G\bm  x+\bm n,
\end{align} 
where $\bm x\in\mathbb{C}^{M_{\rm t}}$ is the signal transmitted from agent A and $\bm n\sim \mathcal{CN}(\bm 0,\sigma_{\rm dl}^2\bm I)$ is the additive white Gaussian noise. 

We design the system to support $N_{\mathrm s}$ independent data streams, so 
the transmit signal $\bm x$ can be represented as $\bm x=\bm W_{\rm t}\bm s$, where $\bm s\in\mathbb{C}^{N_{\rm s}}$ denotes the data symbol vector with $\mathbb{E}[\bm s\bm s^{\sf H}]= \frac{1}{N_{\rm s}}\bm I$ and $\bm W_{\rm t}\in\mathbb{C}^{M_{\rm t}\times N_{\rm s}}$ is the precoding matrix at agent A. We choose the number of data streams to be less than the rank of the channel, i.e., $N_{\rm s}\le \operatorname{rank}(\bm G)$. The transmitted signal must satisfy a normalized power constraint, i.e., $\mathbb{E}[\|\bm x\|_2^2]=\frac{1}{N_{\rm s}}\|\bm W_{\rm t}\|_F^2\le 1$. 

After receiving the signal $\bm y$, agent B applies a combiner matrix $\bm W_{\rm r}\in\mathbb{C}^{M_{\rm r}\times N_{\rm s}}$ to obtain:
\begin{align}\label{eq:sys_model}
  \hat{\bm s} = \bm W_{\rm r}^{\sf H}\bm y= \bm W_{\rm r}^{\sf H} \bm G\bm W_{\rm t}\bm s+\bm W_{\rm r}^{\sf H} \bm n.
\end{align}
The overall achievable rate can be written as  \cite{cover_elements_2006}:
\begin{align}
\label{eq:rate}
  R = \log_2\det(\bm I+\bm C^{-1} \bm W_{\rm r}^{\sf H}\bm G\bm W_{\rm t}\bm W_{\rm t}^{\sf H}\bm G^{\sf H}\bm W_{\rm r}),
\end{align}
where $\bm C=\sigma_{\rm dl}^2\bm W_{\rm r}^{\sf H}\bm W_{\rm r}$. This paper considers the problem of maximizing the achievable rate by optimizing the precoding matrix $\bm W_{\rm t}$ and the combiner matrix $\bm W_{\rm r}$.


If the channel $\bm G$ is known perfectly, the optimization of $\bm W_{\rm t}$ and $\bm W_{\rm r}$ has an analytic solution, namely $\bm W_{\rm t}$ and $\bm W_{\rm r}$ should match the right and left singular vectors of the channel matrix $\bm G$ corresponding to the largest $N_{\rm s}$ singular values. 
Let the SVD of $\bm G$ be 
\begin{align}\label{eq:svd}
  \bm G = [\bm U_1, \bm U_2]\begin{bmatrix}
    &\bm\Sigma_1,&\bm 0\\ &\bm 0, &\bm\Sigma_2
  \end{bmatrix}\begin{bmatrix}
    \bm V_1^{\sf H}\\\bm V_2^{\sf H}
  \end{bmatrix},
\end{align}
where $\bm U_1\in\mathbb{C}^{M_{\rm r}\times N_{\rm s}}$ and $\bm V_1\in\mathbb{C}^{M_{\rm t}\times N_{\rm s}}$ have orthogonal and unit 2-norm columns, and $\bm \Sigma_1$ is a diagonal matrix with top-$N_{\rm s}$ singular values. To maximize the achievable rate (\ref{eq:rate}), an optimal solution $\bm W^\ast_{\rm t}$ and $\bm W^\ast_{\rm r}$ is simply 
\begin{subequations}\label{eq:opt_precoding_decoding}
  \begin{align}
    &\bm W^\ast_{\rm t} = \bm V_1\bm D,\\
    &\bm W^\ast_{\rm r} = \bm U_1,
  \end{align}
\end{subequations}
where $\bm D$ is a diagonal matrix with the diagonal terms given by the water-filling power allocations scheme \cite{cover_elements_2006}. For simplicity, we assume a uniform power allocation scheme for the rest of the paper, i.e., $\bm D=\bm I$, which is near optimal in the high SNR regime. 
Note that the optimal $\bm W_{\mathrm t}$ and $\bm W_{\mathrm r}$ are not unique, as long as the 
subspaces spanned by their columns match that of $\bm V_1$ and $\bm U_1$, respectively.

This paper addresses the case where the channel matrix $\bm G$ is unknown. A conventional technique for designing $\bm W_{\rm t}$ and $\bm W_{\rm r}$ would have required to first estimate the matrix channel in a pilot stage, followed by the SVD of the estimated channel. However, estimating the entire channel matrix would require significant pilot overhead, especially in massive MIMO systems. 

To reduce pilot training overhead, we make a key observation that the optimal
transmit precoding and receive combiner matrices, i.e., \eqref{eq:opt_precoding_decoding},
are only functions of the top-$N_{\rm s}$ singular vectors of $\bm G$, rather than 
the entire channel matrix. Thus, it should be possible to directly learn the transmit
precoding and receive combiner matrices by probing only low-dimensional subspaces 
of the channel matrix. 

This motivates us to propose an active sensing framework consisting of ping-pong pilot
transmissions over multiple stages. Assuming channel reciprocity, 
the two agents send $N_{\mathrm s}$ beamformed pilots in an alternating fashion, thus probing 
the channel in a specific $N_{\mathrm s}$-dimensional subspace in each stage. The idea 
is that each agent would actively design the transmit and receive sensing beamformers 
based on the sequence of observations already made about the channel, so that over the 
multiple
stages the two agents would gradually discover the directions of the top-$N_{\rm s}$ 
singular vectors.

\subsection{Ping-Pong Pilots Transmission}\label{sec:pingpongpilots}
\begin{figure}[t]
 \definecolor{rx}{rgb}{0.0, 0.5, 0.0}
 \centering
  \includegraphics[width=9.5cm]{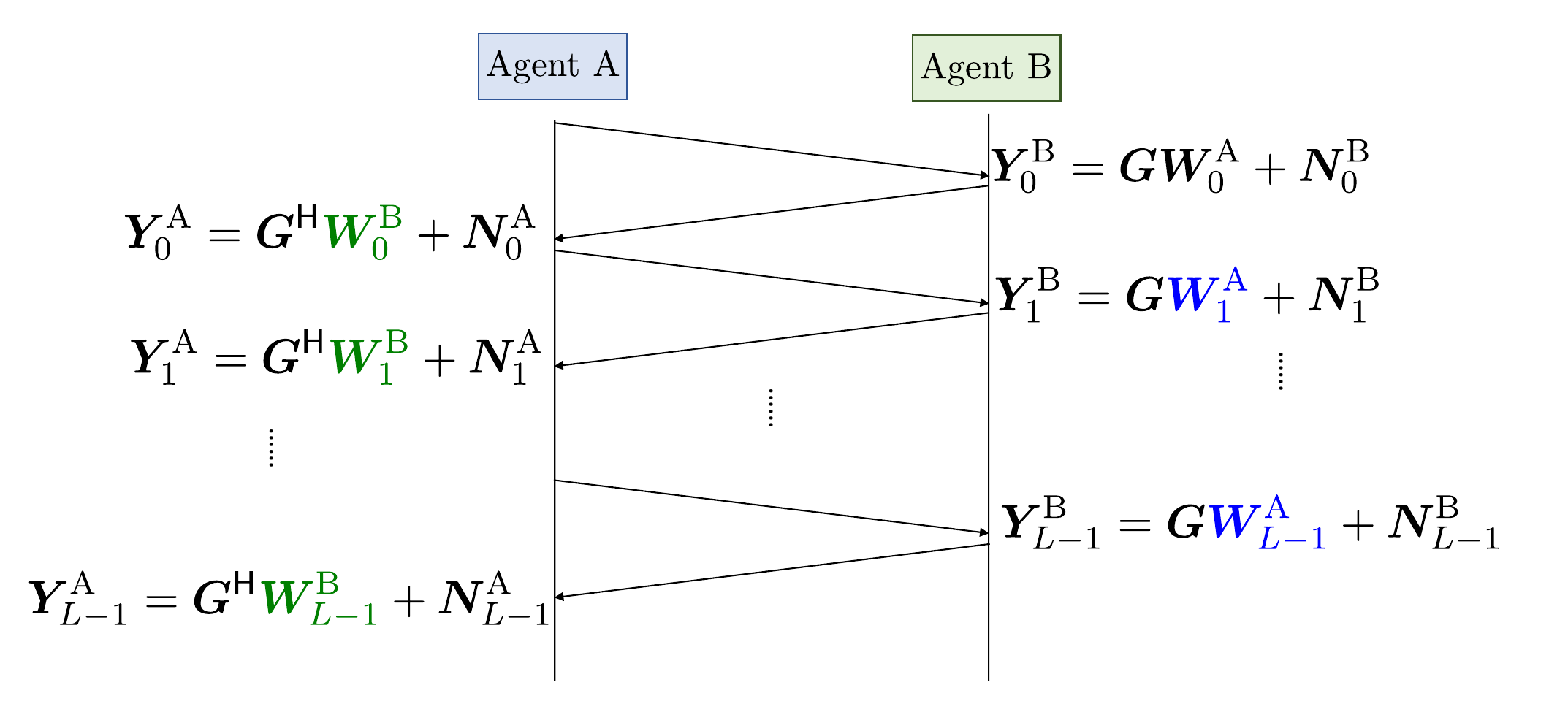}
  \caption{Ping-pong pilot transmission protocol of $L$ rounds for digital MIMO. The sensing matrices adaptively designed at agent A and agent B are highlighted as green (e.g., \textcolor{rx}{$\bm W_{0}^{\rm B}$}) and blue (e.g., \textcolor{blue}{$\bm W_{1}^{\rm A}$}), respectively.}
  \label{fig:pingpong_pilots}
  \vspace{-0.2cm}
\end{figure}
 
The proposed ping-pong pilot transmission protocol is presented in Fig.~\ref{fig:pingpong_pilots}, which consists of $L$ rounds of ping-pong pilot transmission from both sides. In the $\ell$-th round, agent A sends a sequence of pilots $\bm W_{\ell}^{\rm A}\in\mathbb{C}^{M_{\rm t}\times N_{\rm s}}$ of length $N_{\rm s}$, and agent B observes $ \bm Y_{\ell}^{\rm B}$ according to:
\begin{align}\label{eq:pilot_BA}
  \bm Y_{\ell}^{\rm B} = \bm G\bm W_{\ell}^{\rm A}+\bm N_{\ell}^{\rm B},\quad \ell=0,\ldots,L-1,
\end{align}
where $\bm N_{\ell}^{\rm B}$ is the additive white Gaussian noise with each column independently distributed as $\mathcal{CN}(\bm 0,\sigma^{2}_{\rm B}\bm I)$. 
The columns of $\bm W_{\ell}^{\rm A}$, denoted as $\bm w_{\ell,i}^{\rm A}$, are pilot vectors that satisfy the power constraint $\|\bm w_{\ell,i}^{\rm A}\|_2\le 1$. 
After receiving the pilots, agent B sends back a sequence of
pilots $\bm W_{\ell}^{\rm B}\in\mathbb{C}^{M_{\rm r}\times N_{\rm s}} $. Due to
channel reciprocity, agent A receives pilots $ \bm Y_{\ell}^{\rm A}$ as
follows: \begin{align}\label{eq:pilot_AB}
  \bm Y_{\ell}^{\rm A} = \bm G^{\sf H}\bm W_{\ell}^{\rm B}+\bm N_{\ell}^{\rm A},\quad\ell=0,\ldots,L-1,
\end{align}
where $\bm N_{\ell}^{\rm A}$ is the additive white Gaussian noise with each column independently distributed as $\mathcal{CN}(\bm 0,\sigma_{\rm A}^2\bm I)$. The columns of $\bm W_{\ell}^{\rm B}$, denoted by $\bm w_{\ell,i}^{\rm B}$, also satisfy the power constraint $\|\bm w_{\ell,i}^{\rm B}\|^2\le 1$. The pilot vectors $\bm w_{\ell,i}^{\rm A}$ and $\bm w_{\ell,i}^{\rm B}$, $i=1,\cdots,N_{\rm s}$, can be thought of as sensing beamformers because they are used to sense the channel in an $N_{\rm s}$ dimensional subspace. 

In conventional channel estimation approaches, the sensing
beamformers (i.e., pilot sequence) are often generated at random, 
but this is not the most efficient way to probe the low-dimensional part
of the channel matrix $\bm G$ corresponding to its top-$N_{\rm s}$ singular vectors. 
In this paper, we propose to actively design the sensing beamformers based on the
pilots received so far, so that the sensing beamformers learn to focus energy
towards the directions of the top-$N_{\rm s}$ singular vectors.

Specifically, the sensing beamformers $\bm W_{\ell}^{\rm A}$ and $\bm W_{\ell}^{\rm B}$ are designed based on the past received pilots on each side as follows: 
\begin{subequations}\label{eq:sensing}
  \begin{align}
    \bm W_{\ell+1}^{\rm A}= f_{\ell}^{\rm A}([\bm Y_{0}^{\rm A},\ldots,\bm Y_{\ell}^{\rm A}]),\quad\ell=0,\ldots,L-2,\\
    \bm W_{\ell}^{\rm B}= f_{\ell}^{\rm B}([\bm Y_{0}^{\rm B},\ldots,\bm Y_{\ell}^{\rm B}]),\quad\ell=0,\ldots,L-1,
  \end{align}
\end{subequations}
where $f_{\ell}^{\rm A}:\mathbb{C}^{M_{\rm t}\times N_{\rm s}(\ell+1)}\rightarrow \mathbb{C}^{M_{\rm t}\times N_{\rm s}}$ and $f_{\ell}^{\rm B}:\mathbb{C}^{M_{\rm r}\times N_{\rm s}(\ell+1)}\rightarrow \mathbb{C}^{M_{\rm r}\times N_{\rm s}}$ are the sensing strategies of the $\ell$-th transmission round at agent A and agent B, respectively. The columns of $\bm W_{\ell}^{\rm A}$ and $\bm W_{\ell}^{\rm B}$ are the sensing beamformers, which are normalized to unit 2-norm to meet the power constraint.
After $L$ rounds of ping-pong pilot transmission, each agent utilizes all the received pilots on each side to design the data transmission precoding matrix and combining matrix as follows:
\begin{subequations}\label{eq:data}
  \begin{align}
    \bm W_{\rm t}= g^{\rm A}([\bm Y_{0}^{\rm A},\ldots,\bm Y_{L-1}^{\rm A}]),\\
    \bm W_{\rm r}= g^{\rm B}([\bm Y_{0}^{\rm B},\ldots,\bm Y_{L-1}^{\rm B}]),
  \end{align}
\end{subequations}
where $g^{\rm A}:\mathbb{C}^{M_{\rm t}\times N_{\rm s}L}\rightarrow \mathbb{C}^{M_{\rm t}\times N_{\rm s}}$ and $g^{\rm B}:\mathbb{C}^{M_{\rm r}\times N_{\rm s}L}\rightarrow \mathbb{C}^{M_{\rm r}\times N_{\rm s}}$ are functions to map all the received pilots to the desired solutions. Since this paper focuses on learning the directions of singular vectors, a unit 2-norm constraint is enforced to the columns in $\bm W_{\rm t}$ and $\bm W_{\rm r}$, i.e., $\|\bm w_{{\rm t},i}\|^2=1$ and $\|\bm w_{{\rm r},i}\|^2=1$ for all $i$. 

The goal of this paper is to find the mappings $\{f_{\ell}^{\rm A}\}_{\ell=0}^{L-2}$, $\{f_{\ell}^{\rm B}\}_{\ell=0}^{L-1}$, $g^{\rm A}$ and $g^{\rm B}$ such that column space of the final precoding and combining matrices, $\bm W_{\rm t}$ and $\bm W_{\rm r}$, closely match that of $\bm V_1$ and $\bm U_1$, respectively. \textcolor{black}{
  In particular, the optimization problem can be formulated as:
  \begin{equation}\label{eq:formulation1}
    \begin{aligned} 
      &\underset{\mathcal{F},\bm Q_{\rm v},\bm Q_{\rm u}}{\operatorname{minimize}}~&&\mathbb{E}\left[\|{\bm W}_{{\rm t}}-\bm V_1\bm Q_{\rm v} \|_F^2+ \|{\bm W}_{{\rm r}}-\bm U_1\bm Q_{\rm u}\|_F^2 \right]\\
      &\operatorname{subject~to}~&& \eqref{eq:sensing}, \eqref{eq:data},
    \end{aligned}
  \end{equation}
  where the expectation is taken over the channel and noise distributions, 
  the optimization variables are over a set of functions
  $\mathcal{F}=\{\{f_{\ell}^{\rm A}\}_{\ell=0}^{L-2}, \{f_{\ell}^{\rm B}\}_{\ell=0}^{L-1},g^{\rm A},g^{\rm B}\}$, and finally $\bm Q_{\rm v}$ and $\bm Q_{\rm u}$ are  $N_{\rm s}\times N_{\rm s}$ orthogonal matrices to account for the fact that only the subspaces need to match.  
}

Problem \eqref{eq:formulation1} is highly nontrivial because it involves searching for solutions in a high-dimensional functional space. To tackle this issue, we propose a data-driven approach to first parameterize the functions in $\mathcal{F}$ using deep neural networks. We then learn an effective sensing strategy from the training data, consisting of numerous channel realizations based on the communication scenarios of interest. 

Preparing a sufficient amount of training data is crucial in the data-driven approach. However, the objective function defined in  \eqref{eq:formulation1} is not easy to deal with, because 
optimization over $\bm Q_{\rm v}$ and $\bm Q_{\rm u}$ is involved, and further it requires performing SVD for every channel matrix in the training data set in order to obtain $\bm V_1$ and $\bm U_1$.  
To avoid this complexity, we propose to replace the objective function in \eqref{eq:formulation1} by $-\mathbb{E}\left[\log|\det(\bm W_{\rm r}^{\sf H} \bm G\bm W_{\rm t})|^2\right]$, because the function $\log|\det(\bm W_{\rm r}^{\sf H} \bm G\bm W_{\rm t})|^2$ is maximized when ${\bm W}_{{\rm t}}=\bm V_1 $ and ${\bm W}_{{\rm r}}=\bm U_1 $. 
In this way, we reformulate the active sensing problem as:
\begin{equation}\label{eq:formulation2}
  \begin{aligned}
    &\underset{\mathcal{F}}{\operatorname{maximize}}~&&\mathbb{E}\left[\log|\det(\bm W_{\rm r}^{\sf H} \bm G\bm W_{\rm t})|^2\right]\\
    &\operatorname{subject~to}~&& \eqref{eq:sensing}, \eqref{eq:data}.
  \end{aligned}
\end{equation}
Furthermore, we propose to parameterize the functional space $\mathcal{F}$
based on a carefully designed active sensing framework using RNN.  This
parameterization of the functions using RNNs enables us to exploit the
sequential nature of the active sensing problem in order to make more informed
and adaptive sensing decisions.

\section{Proposed Learning-Based Active Sensing Framework for Fully Digital MIMO System}\label{sec:active_unit}
This section presents the proposed active sensing framework for solving problem \eqref{eq:formulation2}. Before delving into the specifics of the framework, we provide an overview of a conventional power iteration method for estimating the dominant singular vectors to motivate the proposed learning-based approach.

\subsection{Motivations From Power Iteration Method}\label{sect:cpit}
Prior work \cite{1323251} proposes a ping-pong based active sensing scheme for
finding the dominant singular vectors of a channel matrix using the power
iteration method.  To illustrate this idea, we consider a specific example of
estimating the top singular vector (i.e., $N_{\rm s}=1$) in the noiseless scenario. 

In the $\ell$-th  ping-pong round, each agent sends a pilot symbol through a sensing beamformer,
which is set to be the received pilot in the previous round. 
This implies that $\bm w_{\ell+1}^{\rm A}=\bm y_{\ell}^{\rm A}$ and $\bm w_{\ell}^{\rm B}=\bm y_{\ell}^{\rm B}$. Starting from a random vector  $\bm w_{0}^{\rm A}$, we obtain the following equations after $\ell$ rounds of transmission:
\begin{subequations}\label{eq:power_t}
  \begin{align}
    \bm y_{\ell}^{\rm A} &= (\bm G^{\sf H}\bm G)^{\ell}\bm w_{0}^{\rm A}=\sum_{i}\sigma_i^{2\ell} \beta_i \bm v_i,\\
    \bm y_{\ell}^{\rm B} &= (\bm G\bm G^{\sf H})^{\ell-1}\bm G\bm w_{0}^{\rm A}=\sum_{i}\sigma_i^{2\ell-1}\beta_i \bm u_i,
  \end{align}
\end{subequations}
where we use the fact that $\bm w_{0}^{\rm A}$ can be expressed as $\bm w_{0}^{\rm A}=\sum_{i=1}^{\operatorname{rank}(\bm G)}\beta_i\bm v_i+ \bar{\bm v}_{{\rm t},0}$, with $\bar{\bm v}_{{\rm t},0}$ belonging to the null space of $\bm G$. As the number of transmission rounds $\ell$ increase, the vectors $\bm y_{\ell}^{\rm A}$ and $\bm y_{\ell}^{\rm B}$ in \eqref{eq:power_t} would be dominated by the top singular vectors $\bm v_1$ and $\bm u_1$, respectively, with a linear convergence rate \cite{1323251}. Furthermore, to meet the transmit power constraint, the unit 2-norm normalization is performed in each iteration, i.e., $\bm w_{\ell}^{\rm B} = {\bm y_{\ell}^{\rm B}}/{\|\bm y_{\ell}^{\rm B}\|_2}$ and $\bm w_{\ell+1}^{\rm A} = {\bm y_{\ell}^{\rm A}}/{\|\bm y_{\ell}^{\rm A}\|_2}$, which does not alter the directions of $\bm y_{\ell}^{\rm A}$ and $\bm y_{\ell}^{\rm B}$. The rapid convergence rate of the power iteration scheme implies a significant pilot overhead reduction as compared to estimating the entire channel matrix. 

The power iteration method can be extended to the multiple data streams ($N_{\rm s}>1$) scenario. Specifically, as presented in \cite{1323251}, agent B performs QR decomposition on the received pilots $\bm Y_{\ell}^{\rm B}$ in \eqref{eq:pilot_BA}, i.e., 
\begin{align}
    \bm Y_{\ell}^{\rm B}=\bm Q^{\rm B}_{\ell}\bm R_{\ell}^{\rm B},
\end{align}
then sends back $N_{s}$ pilots to agent A by setting 
$ \bm W_{\ell}^{\rm B}=\bm Q^{\rm B}_{\ell}$.
Similarly, agent A performs QR decomposition on the received pilots $\bm Y_{\ell}^{\rm A}$ in \eqref{eq:pilot_AB}, i.e., 
\begin{align}
    \bm Y_{\ell}^{\rm A}=\bm Q^{\rm A}_{\ell}\bm R_{\ell}^{\rm A},
\end{align} 
and sets $\bm W_{\ell+1}^{\rm A}=\bm Q^{\rm A}$ for the next transmission. 
The final precoding and combining matrices are set as $\bm W_{\rm t}=\bm W_{L}^{\rm A}$ and $\bm W_{\rm r}=\bm W_{L-1}^{\rm B}$ in the data transmission stage.


The primary impediment to the practical implementation of such a power iteration method is that the algorithm is sensitive to the presence of noise. For this reason, the conventional power iteration method is only suitable for the high SNR scenario.
In this paper, we leverage a data-driven approach to mimic the power iteration method while mitigating its sensitivity to the channel noise by incorporating a neural network based active sensing unit.

\subsection{Proposed Active Sensing Approach}

\begin{figure*}[t]
  \centering
  \includegraphics[width=12cm]{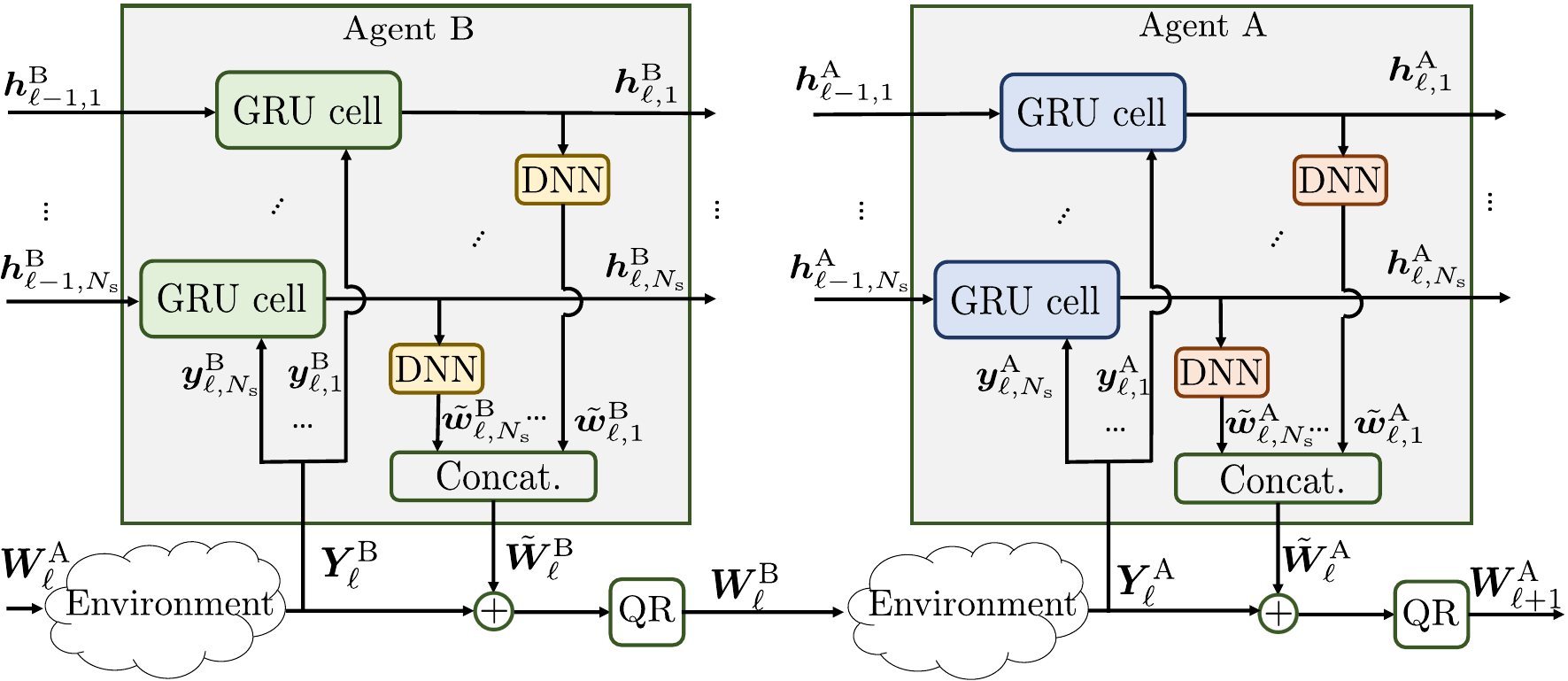}
  \caption{Proposed active sensing unit in the $\ell$-th pilot round for digital MIMO.}
  \label{fig:GRU_unit}
\end{figure*}
\begin{algorithm}[t]
  \begin{algorithmic}[1]
      \STATE Initial $\bm W_{0}^{\rm A}\in\mathbb{C}^{M_{t}\times N_{\rm s}}$, $\bm h_{-1,i}^{\rm B}\in\mathbb{C}^{N_{\rm h}^{\rm B}}$, $\bm h_{-1,i}^{\rm A}\in\mathbb{C}^{N_{\rm h}^{\rm A}}$.
      \FOR{$\ell=0,\ldots,L-1$} 
      \STATE{A$\rightarrow$ B:\quad $\bm Y_{\ell}^{\rm B} = \bm G\bm W_{\ell}^{\rm A}+\bm N_{\ell}^{\rm B}$} 
      \FOR{$i=1,\ldots,N_{\rm s}$}
      \STATE{$\bm h_{\ell,i}^{\rm B}=\operatorname{GRUCell}^{\rm B}(\bm h_{\ell-1,i}^{\rm B},\bm y_{\ell,i}^{\rm B})$}
      \STATE{$\tilde{\bm w}_{\ell,i}^{\rm B} = \operatorname{DNN}^{\rm B}(\bm h_{\ell,i}^{\rm B})$}
      \ENDFOR
      \STATE{$\tilde{\bm W}_{\ell}^{\rm B}=[\tilde{\bm w}_{\ell,1}^{\rm B},\ldots,\tilde{\bm w}_{\ell,N_{\rm s}}^{\rm B}]$}
      \STATE QR decomposition: {($\tilde{\bm W}_{\ell}^{\rm B}+\bm Y_{\ell}^{\rm B}) = \bm Q^{\rm B}_{\ell}\bm R_{\ell}^{\rm B}$}
      \STATE{Set ${\bm W}_{\ell}^{\rm B} = \bm Q^{\rm B}_{\ell}.$}
      \STATE{}
      \STATE{B$\rightarrow$ A:\quad $\bm Y_{\ell}^{\rm A} = \bm G^{\sf H}\bm W_{\ell}^{\rm B}+\bm N_{\ell}^{\rm A}$} 
      \FOR{$i=1,\ldots,N_{\rm s}$}
      \STATE{$\bm h_{\ell,i}^{\rm A}=\operatorname{GRUCell}^{\rm A}(\bm h_{\ell-1,i}^{\rm A},\bm y_{\ell,i}^{\rm A})$}
      \STATE{$\tilde{\bm w}_{\ell,i}^{\rm A} = \operatorname{DNN}^{\rm A}(\bm h_{\ell,i}^{\rm A})$}
      \ENDFOR
      \STATE{$\tilde{\bm W}_{\ell}^{\rm A}=[\tilde{\bm w}_{\ell,1}^{\rm A},\ldots,\tilde{\bm w}_{\ell,N_{\rm s}}^{\rm A}]$}
      \STATE QR decomposition: {($\tilde{\bm W}_{\ell}^{\rm A}+\bm Y_{\ell}^{\rm A}) = \bm Q^{\rm A}_{\ell}\bm R_{\ell}^{\rm A}$}
      \STATE{Set ${\bm W}_{\ell+1}^{\rm A} = \bm Q^{\rm A}_{\ell}$}
      \ENDFOR
  \end{algorithmic}
  \caption{Proposed active sensing framework for digital MIMO.}\label{algo1}
\end{algorithm}

We now describe the proposed active sensing framework for learning the sensing and data transmission matrices. The proposed framework is designed to be robust to channel noise and can achieve excellent performance even in low SNR scenarios. Moreover, the proposed framework includes the power iteration method as a special case.

The proposed active sensing framework parameterizes the sensing strategies $f_{\ell}^{\rm A}(\cdot)$ and $f_{\ell}^{\rm B}(\cdot)$ using an active sensing unit as shown in Fig.~\ref{fig:GRU_unit}. In the active sensing unit, both agents A and B use $N_{\rm s}$ gated recurrent unit (GRU) cells to extract useful information from the received pilots and $N_{\rm s}$ DNNs to map the latest hidden state vectors to the next sensing vectors. More specifically, given the observation $\bm Y_{\ell}^{\rm B}$ in \eqref{eq:pilot_BA}, agent B takes the $i$-th column of $\bm Y_{\ell}^{\rm B}$ (denoted by $\bm y_{\ell,i}^{\rm B}$) as input to the $i$-th GRU cell. The GRU cell updates its hidden state vector $\bm h_{\ell,i}^{\rm B}\in\mathbb{C}^{N_{\rm h}^{\rm B}}$ as follows:
\begin{align}\label{eq:update_hidden}
\bm h_{\ell,i}^{\rm B}=\operatorname{GRUCell}^{\rm B}(\bm h_{\ell-1,i}^{\rm B},\bm y_{\ell,i}^{\rm B}),\quad i=1,\ldots,N_{\rm s}.
\end{align}
Here $\operatorname{GRUCell}^{\rm B}(\bm h_{\ell-1,i}^{\rm B},\bm y_{\ell,i}^{\rm B})$ denotes the standard GRU implementation as follows \cite{cho2014properties}:
\begin{subequations}\label{eq:gru_imp}
\begin{align}
&\bm r = \sigma(\bm W_{\rm ir} \tilde{\bm y}_{\ell,i}^{\rm B} +\bm b_{\rm ir} + \bm W_{\rm hr} \bm h_{\ell-1,i}^{\rm B} + \bm b_{\rm hr}), \\
&\bm z = \sigma(\bm W_{\rm iz} \tilde{\bm y}_{\ell,i}^{\rm B} + \bm b_{\rm iz} + \bm W_{\rm hz} \bm h_{\ell-1,i}^{\rm B} + \bm b_{\rm hz}), \\
&\bm n = \tanh(\bm W_{\rm in} \tilde{\bm y}_{\ell,i}^{\rm B} + \bm b_{\rm in} + \bm r \circ (\bm W_{\rm hn} \bm h_{\ell-1,i}^{\rm B} +\bm b_{\rm hn})), \\
&\bm h_{\ell,i}^{\rm B} = (1 - \bm z) \circ \bm n + \bm z \circ \bm h_{\ell-1,i}^{\rm B},
\end{align}
\end{subequations}
where $\tilde{\bm y}_{\ell,i}^{\rm B} = [\Re{({\bm y}_{\ell,i}^{\rm B})}^\top, \Im{({\bm y}_{\ell,i}^{\rm B})}^\top ]^\top$, $\circ$ is the Hadamard product, and $\sigma(\cdot)$ is the element-wise sigmoid function. The trainable weights and bias in the GRU are $\{\bm W_{\rm ir}, \bm W_{\rm hr}, \bm W_{\rm iz},\bm W_{\rm hz},\bm W_{\rm in},\bm W_{\rm hn} \}$ and $\{\bm b_{\rm ir},\bm b_{\rm hr},\bm b_{\rm iz},\bm b_{\rm hz},\bm b_{\rm in},\bm b_{\rm hn} \}$.
\textcolor{black}{The set of trainable parameters within the GRU cell is shared across all the ping-pong rounds within each agent to reduce training complexity, but the two agents have different trainable parameters.}

The ability of GRU to summarize useful information from historical observations into a vector of fixed dimension enables the proposed active sensing scheme to scale up to any number of transmission rounds, since the same unit in Fig.~\ref{fig:GRU_unit} can be applied to all transmission rounds. We remark that other implementations of RNN such as long short-term memory network (LSTM) can also be used here, the choice of GRU is due to its memory efficiency.

The hidden state vector $\bm h_{\ell,i}^{\rm B}$ is then mapped to a vector $\tilde{\bm w}_{\ell,i}^{\rm B}\in\mathbb{C}^{M_{\rm t}}$ using a fully connected neural network $\operatorname{DNN}^{\rm B}$:
\begin{align}\label{eq:tilde_w}
\tilde{\bm w}_{\ell,i}^{\rm B} = \operatorname{DNN}^{\rm B}(\bm h_{\ell,i}^{\rm B}),\quad i=1,\ldots,N_{\rm s}.
\end{align}
The vectors $\tilde{\bm w}_{\ell,i}^{\rm B}$'s are collected into a matrix $\tilde{\bm W}_{\ell}^{\rm B}$, given by
\begin{align}\label{eq:cat_w}
\tilde{\bm W}_{\ell}^{\rm B}=[\tilde{\bm w}_{\ell,1}^{\rm B},\ldots,\tilde{\bm w}_{\ell,N_{\rm s}}^{\rm B}].
\end{align}
After power normalization in each column, the matrix $\tilde{\bm W}_{\ell}^{\rm B}$ can already be used as the designed sensing beamformers in the $\ell$-th transmission round at agent B.

\textcolor{black}{To further improve the performance, we propose to incorporate elements of the power iteration method into the neural network through the following steps:
\begin{subequations}\label{eq:final_W}
  \begin{align}
    &(\tilde{\bm W}_{\ell}^{\rm B}+\bm Y_{\ell}^{\rm B}) = \bm Q^{\rm B}_{\ell}\bm R_{\ell}^{\rm B} \quad(\text{QR decomposition}),\label{eq:rsnet}\\
    &{\bm W}_{\ell}^{\rm B} = \bm Q^{\rm B}_{\ell}.
  \end{align}
\end{subequations}
This ensures that the proposed active sensing framework performs at least as well as the conventional power method. If the neural network generates a matrix $\tilde{\bm W}_{\ell}^{\rm B}=\bm 0$,  the proposed active sensing framework reduces to the conventional power method proposed in \cite{1323251}. Moreover, we observe in simulations that the neural network is easier to train with the step \eqref{eq:rsnet} of adding the last observation $\bm Y_{\ell}^B$. This approach bears resemblance to the structure of residual neural networks (ResNet) \cite{he2016deep}, where the incorporation of the last observation \(\bm Y_{\ell}^B\) can be considered akin to a residual connection. Such connections have been shown to facilitate gradient backpropagation during the training phase \cite{he2016deep}.}

The neural network architecture at agent A is the same as agent B, except that it uses different parameters. The overall algorithm in the ping-pong pilot transmission stage is listed in Algorithm~\ref{algo1}. 
The initial matrix $\bm W_{0}^{\rm A}\in\mathbb{C}^{M_{\rm t}\times N_{\rm s}}$ is learned from channel statistics in the training stage\footnote{\textcolor{black}{This can be achieved by setting \(\bm W_0^{\rm A}\) as a trainable parameter in the training stage, allowing it to be automatically updated through the gradient backpropagation process.}} and remains fixed in the testing stage.  The initial hidden state vectors $\bm h_{-1,i}^{\rm B}\in\mathbb{C}^{N_{\rm h}^{\rm B}}$ and $\bm h_{-1,i}^{\rm A}\in\mathbb{C}^{N_{\rm h}^{\rm A}}$ are both set to be the vector with all entries equal to one. To improve scalability, each side has $N_{\rm s}$ copies of the GRU cells and DNNs to design the sensing beamformers for the $N_{\rm s}$ different data streams.
 
To design the final precoding and combining matrices in the data transmission stage, both agents use another set of DNNs to map the latest hidden state vectors to the corresponding matrices, which are further processed using QR decomposition. After $L$ rounds of pilot transmission, agent B generates the final combining matrix ${\bm W}_{{\rm r}}$ as follows:
\begin{subequations}\label{eq:qr_r}
  \begin{align}
    &\tilde{\bm w}_{{\rm r},i} =\operatorname{DNN}^{\rm r}(\bm h_{L,i}^{\rm B}),\quad i=1,\ldots,N_{\rm s},\label{eq:dnn_r}\\
    &\tilde{\bm W}_{{\rm r}} = [\tilde{\bm w}_{{\rm r},1},\ldots,\tilde{\bm w}_{{\rm r},N_{\rm s}}],\\
    &(\tilde{\bm W}_{{\rm r}}+\bm Y_{L-1}^{\rm B}) = \bm Q^{\rm r}\bm R^{\rm r},\quad(\text{QR decomposition}),\\
    &{\bm W}_{{\rm r}} = \bm Q^{\rm r}.
  \end{align}
\end{subequations}
Similarly, agent A uses its hidden state vector $\bm h_{L,i}^{\rm A}$ and received pilots $ \bm Y_{L-1}^{\rm A}$ to design its precoding matrix $\bm W_{\rm t}$ following the same procedure as in \eqref{eq:qr_r}, but replaces the $\operatorname{DNN}^{\rm r}$ and $\bm Y_{L-1}^{\rm B} $ with $\operatorname{DNN}^{\rm t}$ and $\bm Y_{L-1}^{\rm A} $, respectively.

The proposed active sensing framework is trained by sequentially concatenating $L$ active sensing units to learn the global policy for the entire sensing trajectory during the training stage. The loss function is set as $-\mathbb{E}\left[\log|\det(\bm W_{\rm r}^{\sf H} \bm G\bm W_{\rm t})|^2\right]$. Furthermore, to train a neural network to be generalizable to different number of ping-pong rounds $L$, we can set the loss function as $-\mathbb{E}\left[\sum_{\ell=0}^{L-1}\log|\det(\bm W_{{\rm r},\ell}^{\sf H} \bm G\bm W_{{\rm t},\ell})|^2\right]$, where $\bm W_{{\rm r},\ell}$ and $\bm W_{{\rm t},\ell}$ are the data transmission precoding and combining matrices generated by the neural network in round $\ell$.

\section{Performance Evaluation for Fully Digital MIMO Systems}\label{sec:simualtion_digital}
In this section, we evaluate the performance of the proposed active sensing framework for fully digital MIMO systems. \textcolor{black}{The dimensions of the hidden states of the GRUs are set to be $N_{\rm h}^{\rm A} = N_{\rm h}^{\rm B}=512$. All the DNNs are two-layer fully connected neural networks with the dimension $[512,1024,2M_{\rm t}]$ and $\operatorname{relu}(\cdot)$ as the activation function in the hidden layers. The entire neural network involving $L$ concatenated active sensing units (as shown in Fig.~\ref{fig:GRU_unit}) is implemented on PyTorch \cite{Paszke2019PyTorch}, which can automatically compute the gradient of the loss function with respect to the trainable parameters. We train the neural network offline using the Adam optimizer \cite{kingma2015adam}, with a default configuration in PyTorch and with an initial learning rate progressively decreasing from $10^{-3}$ to $10^{-5}$. In each training iteration, the batch size of training data is $1024$. We generate as much training data as needed and stop training if the performance on the validation dataset does not improve after several epochs. The testing dataset consists of $1000$ randomly generated channel matrices. In the training stage, we use the loss function $-\mathbb{E}\left[\sum_{\ell=0}^{L-1}\log|\det(\bm W_{{\rm r},\ell}^{\sf H} \bm G\bm W_{{\rm t},\ell})|^2\right]$ to train a single model that is generalizable to different number of ping-pong rounds.}

The proposed active sensing approach is compared with the following benchmarks:

\textit{LMMSE+SVD}: Agent A sends $2LN_{\rm s}$ pilots to agent B using random sensing beamformers. Agent B first estimates the channel matrix based on the received pilots using a linear minimum mean square error (LMMSE) estimator, then performs SVD on the estimated channel matrix to design precoding and combining matrices of the data transmission stage. Finally, agent B sends the designed precoding matrix to agent A via an error-free feedback channel.

\textit{Power Iteration Method \cite{1323251}:} Two agents send pilots in a ping-pong manner and design the sensing beamformers based on the QR decomposition of the received pilots. The algorithm is described in Section \ref{sect:cpit}.

\textit{Summed Power Method \cite{7947217}:} Two agents send pilots according to the ping-pong protocol. Both agents calculate their next sensing beamformers based on the accumulated sum of previously received pilots to effectively average out noise. 
\label{sec:performance_digital_mimo}
\begin{figure*}
     \centering
     \subfloat[SNR=$-10$dB, $N_{\rm s}=2$]{\includegraphics[width=0.24\textwidth]{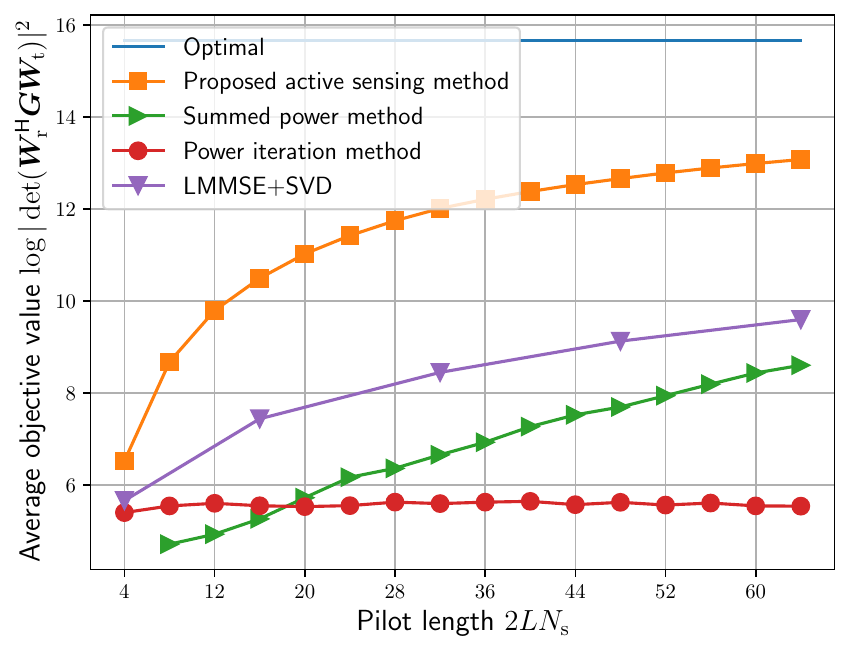}\label{fig:ns2_-10dB}}
     \subfloat[SNR=$-5$dB, $N_{\rm s}=2$]{\includegraphics[width=0.24\textwidth]{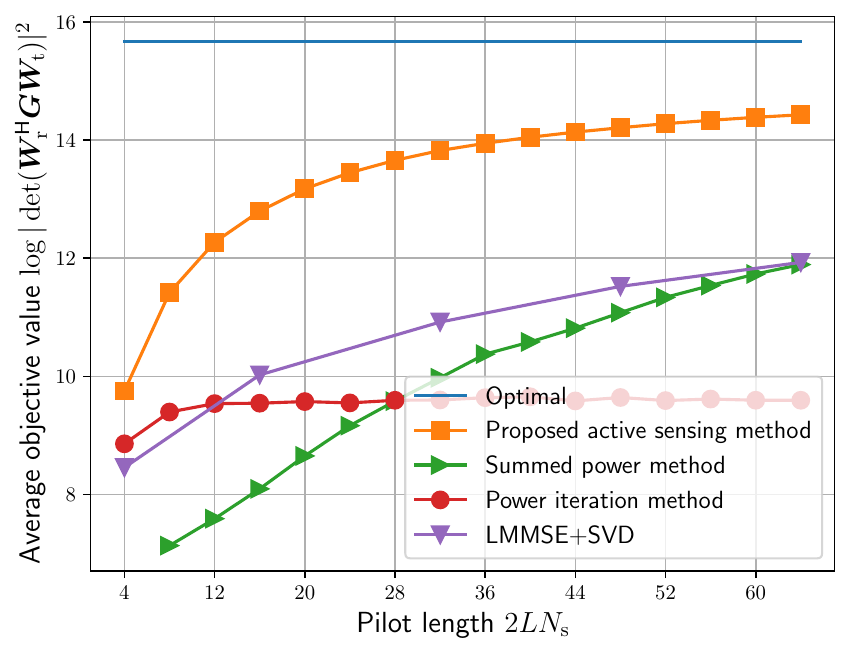}\label{fig:ns2_-5dB}}
     \subfloat[SNR=$0$dB, $N_{\rm s}=2$]{\includegraphics[width=0.24\textwidth]{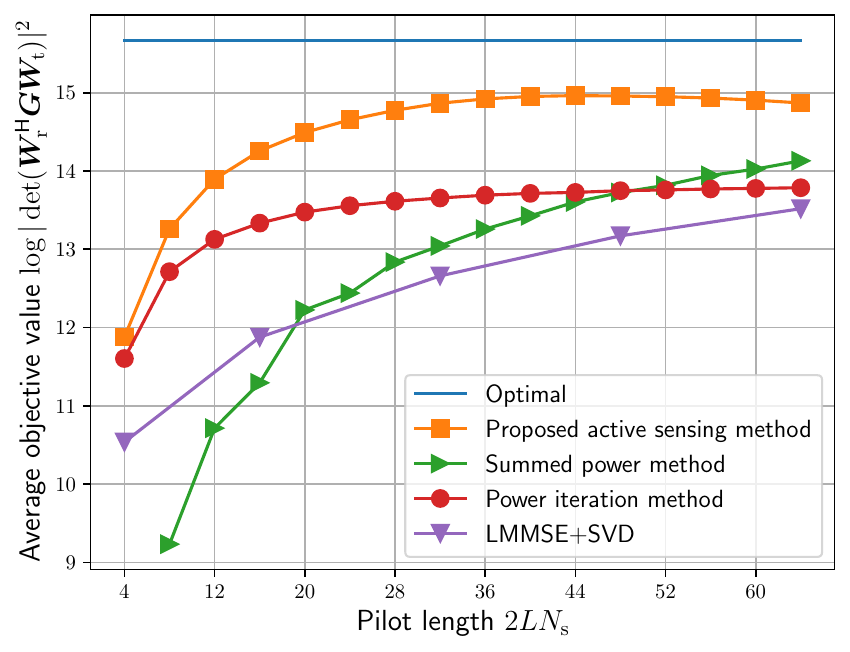}\label{fig:ns2_0dB}}
     \subfloat[SNR=$5$dB, $N_{\rm s}=2$]{\includegraphics[width=0.24\textwidth]{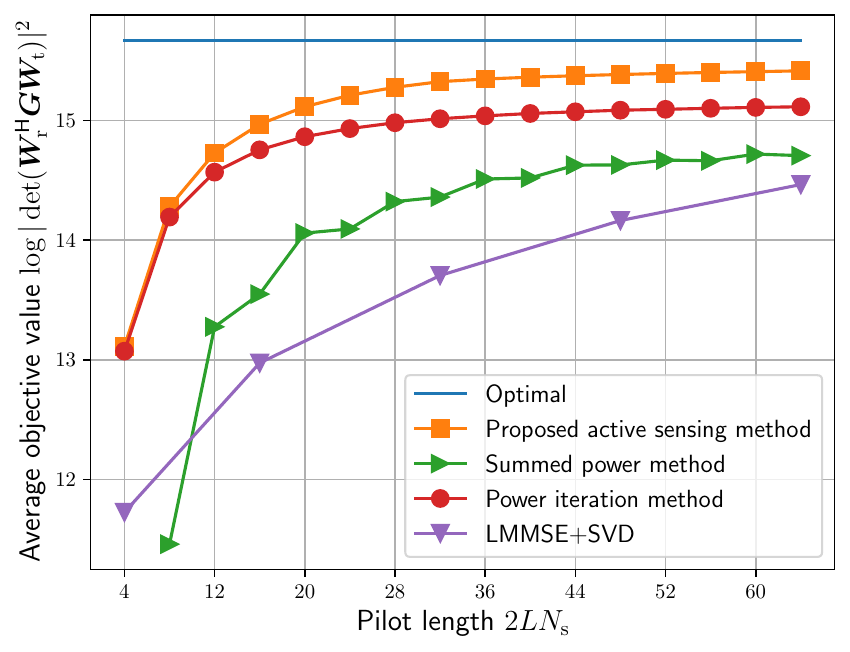}\label{fig:ns2_5dB}}\\
     \subfloat[SNR=$-10$dB, $N_{\rm s}=4$]{\includegraphics[width=0.24\textwidth]{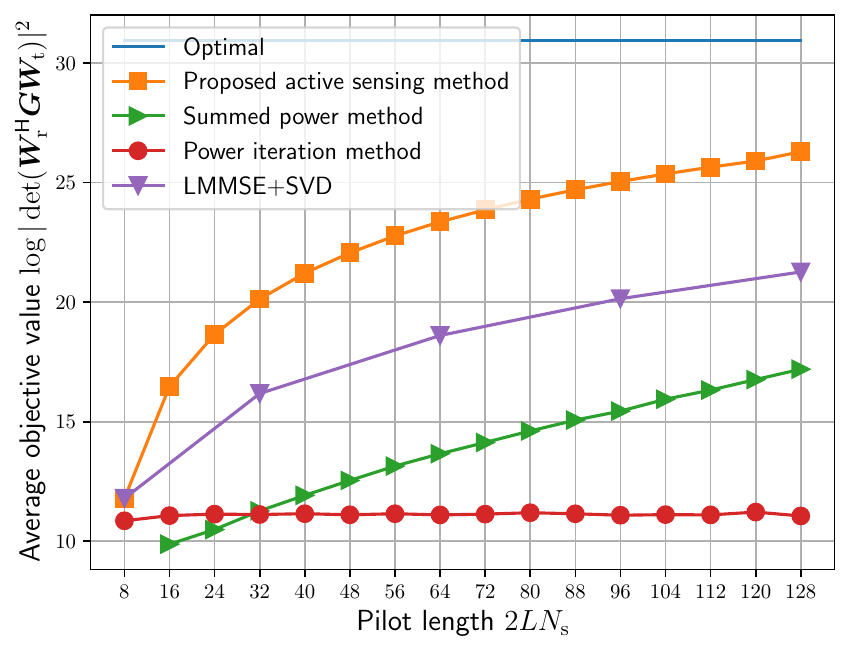}\label{fig:ns4_-10dB}}
     \subfloat[SNR=$-5$dB, $N_{\rm s}=4$]{\includegraphics[width=0.24\textwidth]{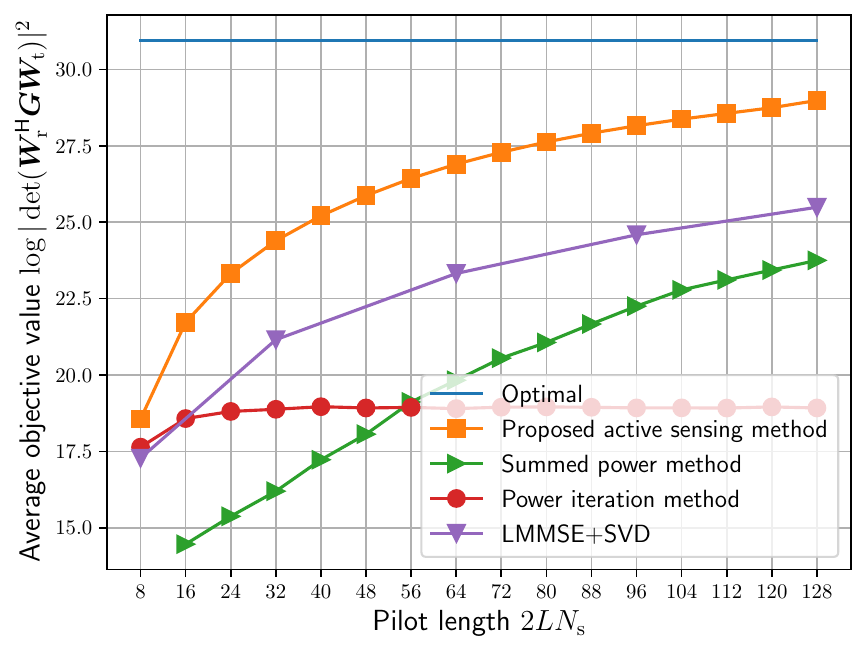}\label{fig:ns4_-5dB}}
     \subfloat[SNR=$0$dB, $N_{\rm s}=4$]{\includegraphics[width=0.24\textwidth]{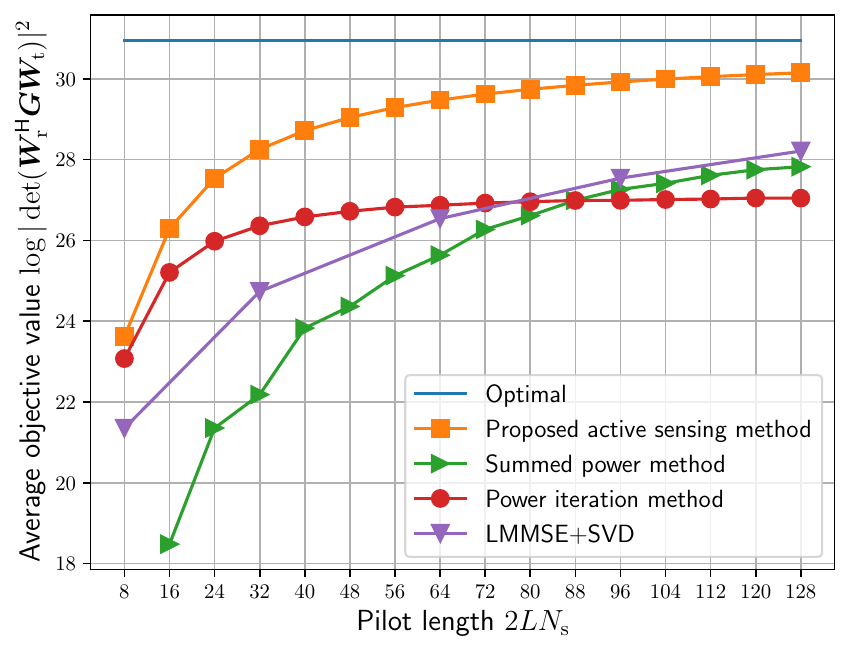}\label{fig:ns4_0dB}}
     \subfloat[SNR=$5$dB, $N_{\rm s}=4$]{\includegraphics[width=0.24\textwidth]{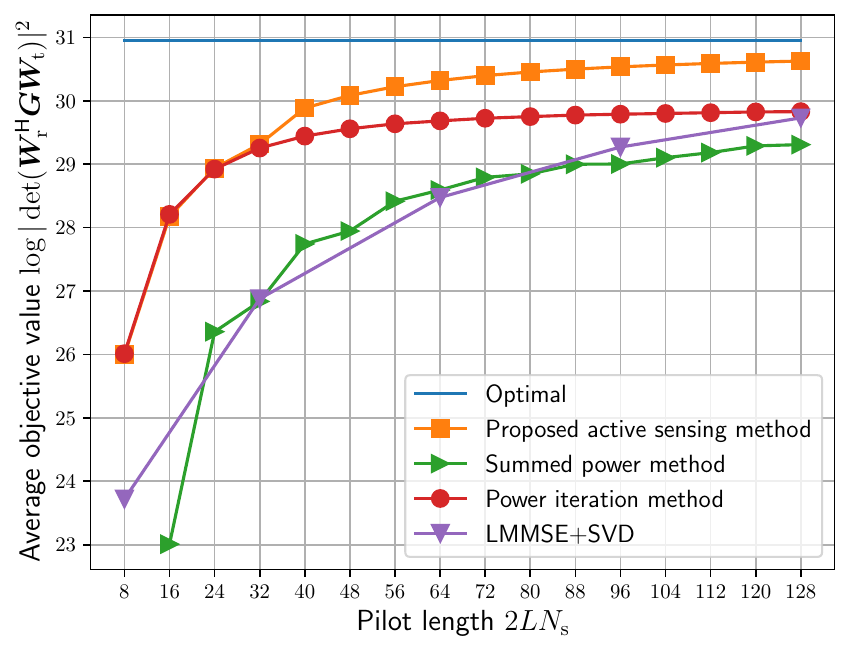}\label{fig:ns4_5dB}}
     \caption{Performance comparison under the  Rayleigh fading channel model for a fully digital MIMO system with $M_{\rm t}=M_{\rm r}=64 $.}
     \label{fig:mimo_3}
\end{figure*}

\begin{figure}[t]
  \centering
  \includegraphics[width=6cm]{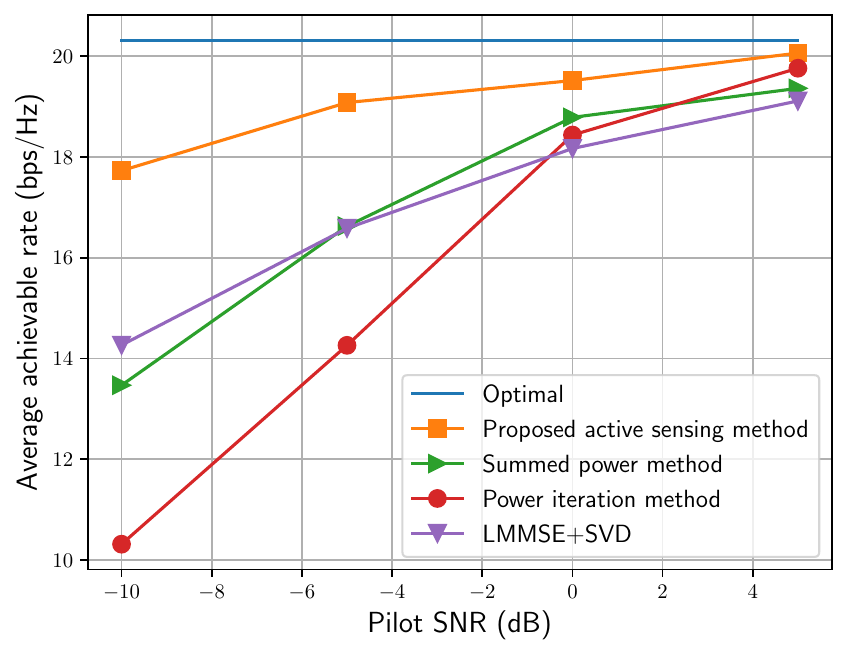}
  \caption{Achievable rate vs. pilot SNR.}
  \label{fig:capacity}
\end{figure}

\begin{figure*}[t]
  \centering
  \subfloat[Ray-tracing model.]{\includegraphics[width=0.4\textwidth]{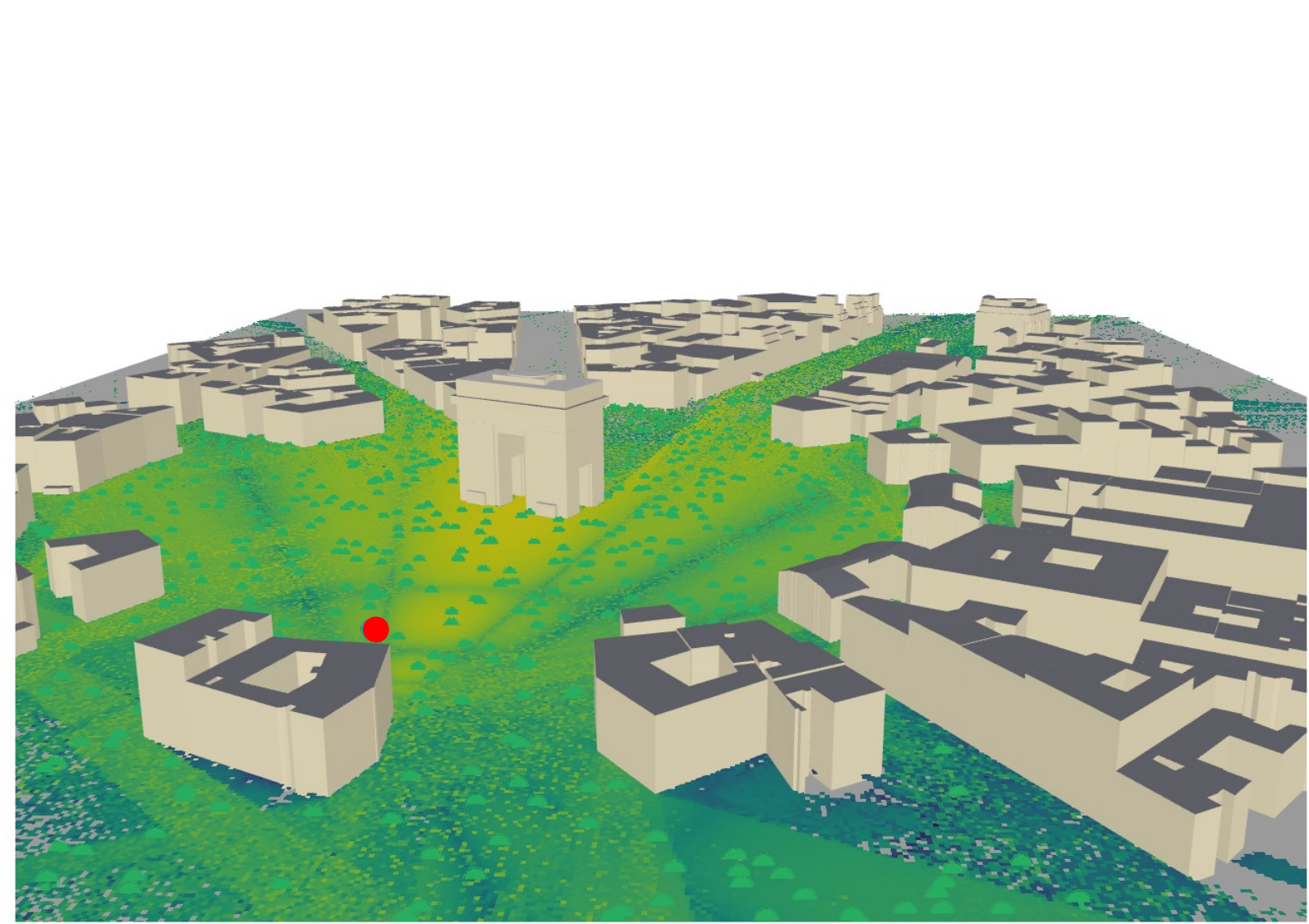} \label{fig:raytracing_scenario}}\hspace{0.8cm}
  \subfloat[Performance comparison.]{\includegraphics[width=0.4\textwidth]{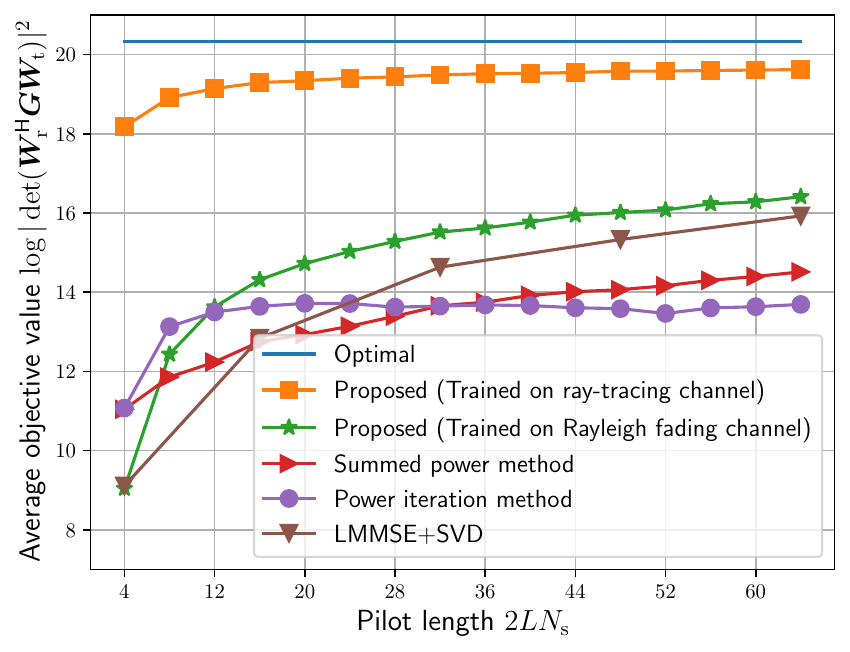} \label{fig:raytracing_results}}
  \caption{Performance comparison under a ray-tracing channel model for a fully digital MIMO system with  $M_{\rm t}=64, M_{\rm r}=16 $ and $N_{\rm s}=2$. }
  \label{fig:mimo_raytracing}
\end{figure*}

\subsection{Rayleigh Fading Channel}
Consider first a MIMO system  with $M_{\rm t}=64$ transmit antennas and $M_{\rm r}=64$ receive antennas. The channel is assumed to follow Rayleigh fading; namely, the entries of $\bm G$ follow i.i.d. Gaussian distribution $\mathcal{CN}(0,1)$. We note that the model should be trained using the site-specific channel data to achieve the best performance. For simplicity, we assume the SNRs in both the uplink and the downlink are the same, i.e., $\text{SNR}\triangleq 1/{\sigma_{\rm A}^2}=1/{\sigma_{\rm B}^2}$.

In Fig.~\ref{fig:mimo_3}, we compare the performances of different methods in terms of the average objective value for varying SNRs and pilot lengths. The results of the scenario with $N_{\rm s}=2$ and $N_{\rm s}=4$ data streams are shown in Fig.~\ref{fig:ns2_-10dB}--\ref{fig:ns2_5dB} and Fig.~\ref{fig:ns4_-10dB}--\ref{fig:ns4_5dB}, respectively. In both scenarios, the neural network is trained for the case where the number of ping-pong rounds is $L=8$ (corresponding to pilot length $2LN_{\rm s}=32$ for $N_{\rm s}=2$ and $2LN_{\rm s}=64$ for $N_{\rm s}=4$).

It can be seen from Fig.~\ref{fig:ns2_-10dB}--\ref{fig:ns2_5dB} that the proposed active sensing method outperforms other benchmarks significantly across different SNRs. For example, in Fig.~\ref{fig:ns2_0dB} and \ref{fig:ns2_5dB}, we can see that the power iteration method and the summed power method converge quickly, but they yield suboptimal solutions due to the impact of noise. At a relatively high SNR $5$ dB, we see some marginal performance gain of the proposed active sensing method over the power iteration and summed power methods. However, the performance gain becomes significant as the SNR decreases, e.g., in Fig.~\ref{fig:ns2_-10dB} and \ref{fig:ns2_-5dB}, from which we observe that the proposed active sensing approach can still achieve good performance, but the power iteration method does not work when the SNR drops to $-10\text{dB}$ or $-5\text{dB}$. This is because the power iteration method is not designed to handle the noise. These observations indicate that the proposed approach is much more robust to noise than the other benchmarks.

\textcolor{black}{From Fig.~\ref{fig:ns2_0dB} and \ref{fig:ns2_5dB}, we observe that the channel estimation based method achieves much worse performance than the proposed active sensing method and is not as competitive as the other benchmarks when $N_{\rm s}=2$, even though the LMMSE estimator is optimal in terms of the mean squared error metric for the Rayleigh fading channel. This performance gap is evident even in the first ping-pong round ($L=1$),  where the initial sensing beamformers are not actively designed since we do not have received pilots yet. This implies that estimating the entire channel matrix is an inefficient scheme if we are only interested in the subspace corresponding to the top-$N_{\rm s}$ singular vectors of the channel matrix.}We also observe that the channel estimation based method achieves better performance than the power iteration method and summed power method when the SNR is low, e.g., $-10$dB and $-5$dB, but its performance is significantly worse than the proposed active sensing approach.

Moreover, we observe that the neural network generalizes well in terms of different numbers of ping-pong rounds $L$. For example, in Fig.~\ref{fig:ns2_-10dB} and~\ref{fig:ns2_-5dB}, the neural network for $N_{\rm s}=2$ is trained when $L=8$, i.e., $2LN_{\rm s}=32$, it can be seen that the performance of the proposed active sensing method keeps improving when the pilot length $2LN_{\rm s}$ is greater than $32$. Similar phenomena can also be observed when $N_{\rm s}=4$.

In Fig.~\ref{fig:ns4_-10dB}--\ref{fig:ns4_5dB}, we further compare the performances of different methods for the scenario $N_{\rm s}=4$. It is observed that the proposed active sensing method still achieves the best performance. However, by comparing Fig.~\ref{fig:ns2_0dB} and Fig.~\ref{fig:ns4_0dB}, we can see that the conventional power iteration method and summed power method is not as competitive as the channel estimation based approach when the number of data streams increases. We should also note that the conventional channel estimation based approach requires an additional error-free feedback channel to transmit the designed precoding matrix from agent B to A. The feedback channel may not be easy to realize when the SNR is low. The other approaches do not need the feedback channel since the two agents can implicitly coordinate at the ping-pong pilot transmission stage. 

\textcolor{black}{In Fig.~\ref{fig:capacity}, we further present the theoretical achievable rate (without employing water-filling) with \(N_{\rm s}=2\), \(M_{\rm r}=64\) and \(M_{\rm t}=64\), where the data transmission SNR is set at $10$dB. From Fig.~\ref{fig:capacity}, it is evident that the proposed active sensing method outperforms other approaches across various pilot SNRs. This figure shows that optimizing the proposed objective function \(\log|\det(\bm{W}_{\rm r}^{\sf H}\bm{G} \bm{W}_{\rm t})|^2\) also approximately maximizes the achievable rate. }

\subsection{Ray-Tracing Based Channel Model} 

\textcolor{black}{We have so far evaluated the proposed active sensing approach under the Rayleigh fading channel model, which is the most challenging scenario for the learning based approach since the neural network cannot exploit any underlying structure of the channel. To evaluate performance in more realistic scenarios, we test the proposed scheme in a ray-tracing based channel model generated from Sionna \cite{sionna}. This scenario, illustrated in Fig.~\ref{fig:raytracing_scenario}, includes a transmitter with $64$ fully digital antennas located at the red point, and a user equipped with $16$ fully digital antennas, randomly positioned within the signal coverage area. The system operates at a carrier frequency of $3.5$ GHz, with a transmit power of $20$ dBm and a noise power of $-114$ dBm during the pilot phase. The results, as depicted in Fig.~\ref{fig:raytracing_results}, demonstrate a remarkable performance improvement of the proposed active sensing method over the other benchmarks. It is noteworthy that the dynamic range for channel strength variation can be almost $60$ dB, indicating that the neural network can be adapted to a broad range of SNRs. Moreover, we also plot the performance of the proposed active sensing method if the neural network is trained on the Rayleigh fading model, which has a completely different characteristics as the testing data. Interestingly, we still observe better performance for the proposed scheme as compared to the other benchmarks. This demonstrates the strong generalization capability of the proposed active sensing methods across different channel distributions.}

\textcolor{black}{We note that the learning-based approach requires training with site-specific data to achieve the best performance. Therefore, for practical deployment, we suggest training different models for various system parameters, such as the SNR, channel statistics, the number of antennas, the number of RF chains, and the array architecture, etc. These models can be stored on site and can be selected based on the current operating scenario. We emphasize that all these models are trained offline, so training complexity is not a concern.}

\section{Active Sensing for Hybrid MIMO System}\label{sec:model_hybrid}
In this section, we extend the proposed active sensing framework to the scenario where both the transmitter and receiver have a hybrid analog and digital beamforming architecture. Hybrid beamforming substantially decreases power consumption as compared to fully digital MIMO systems. However, it also makes channel estimation more challenging, because the hybrid beamforming structure only allows low-dimension observations of the high-dimensional channel. Moreover, the hybrid architecture introduces additional nonconvex analog beamformer constraints. This section aims to modify the proposed active sensing framework to accommodate these new considerations.

\subsection{System Model}
We consider a point-to-point MIMO system with a hybrid antenna architecture. Specifically, agent A with $M_{\rm t}$ antennas and $N_{\rm t}^{\rm RF}$ RF chains sends  $N_{\rm s}$ independent data streams to agent B, which is equipped with $M_{\rm r}$ antennas and $N_{\rm r}^{\rm RF}$ RF chains. Further, it is assumed that $N_{\rm s}\le\min(N_{\rm t}^{\rm RF},N_{\rm r}^{\rm RF})$ and the rank of the channel is greater than $N_{\rm s}$. In the hybrid beamforming structure, the overall precoding matrix $\bm W_{\rm t}\in\mathbb{C}^{M_{\rm t}\times N_{\rm s}}$ at agent A is given by 
\begin{align}\label{eq:w_t}
  \bm W_{\rm t} = \bm F_{\rm t}\tilde{\bm W}_{\rm t},
\end{align}
where $\tilde{\bm W}_{\rm t}\in\mathbb{C}^{N_{\rm t}^{\rm RF}\times N_{\rm s}}$ is the digital precoding matrix and $\bm F_{\rm t}\in\mathbb{C}^{N_{\rm t}\times N_{\rm t}^{\rm RF}}$ is the analog precoding matrix. Since the analog precoder is usually implemented by phase shifters, each element in the analog precoding matrix must satisfy a unit modulus constraint, i.e., $|[\bm F_{\rm t}]_{ij}|=1$, $\forall~i,j$. The overall precoding matrix also needs to satisfy a normalized power constraint $\|\bm W_{\rm t}\|_F^2=\|\bm F_{\rm t}\tilde{\bm W}_{\rm t}\|_F^2\le N_{\rm s}$. At agent B, the hybrid combining matrix $\bm W_{\rm r} $ has a similar structure as follows:
\begin{align}
  \bm W_{\rm r} = \bm F_{\rm r}\tilde{\bm W}_{\rm r},
\end{align}
where $\tilde{\bm W}_{\rm r}\in\mathbb{C}^{N_{\rm r}^{\rm RF}\times N_{\rm s}}$ is the digital combining matrix and $\bm F_{\rm r}\in\mathbb{C}^{N_{\rm r}\times N_{\rm r}^{\rm RF}}$ is the analog combining matrix with the unit modulus constraint $|[\bm F_{\rm r}]_{ij}|=1$, $\forall~i,j$. Therefore, the input and output relationship of the MIMO system is given by
\begin{align}\label{eq:sys_model_hybrid}
  \hat{\bm s} = \tilde{\bm W}_{\rm r}^{\sf H} \bm F_{\rm r}^{\sf H} \bm G\bm F_{\rm t} \tilde{\bm W}_{\rm t}\bm s+\tilde{\bm W}_{\rm r}^{\sf H} \bm F_{\rm t}^{\sf H} \bm n,
\end{align}
where $\bm s\in\mathbb{C}^{N_{\rm s}}$ and $\bm n\in\mathbb{C}^{N_{\rm r}}$ are the transmitted signal and noise, respectively, as defined in \eqref{eq:sys_model}. 

Unlike the fully digital scenario, the problem of finding the optimal hybrid beamforming matrices does not have an analytical solution and is more challenging even when the channel matrix $\bm G$ is perfectly known. For the fully digital MIMO system, the optimal overall precoding matrix $\bm W_{\rm t}^\ast $ and overall combining matrix $\bm W_{\rm r}^\ast$ can be found as in \eqref{eq:opt_precoding_decoding}, but it is not clear whether a further matrix decomposition step on $\bm W_{\rm t}^\ast $ and $\bm W_{\rm r}^\ast$ can be performed to obtain the optimal analog and digital beamformers. One approach for finding the hybrid precoding beamformers is to formulate the following matrix decomposition problem \cite{7397861}:
\begin{equation}\label{eq:decomp1}
  \centering
  \begin{aligned}
    &\underset{\bm F_{\rm t}, \tilde{\bm W}_{\rm t}}{\operatorname{minimize}}~&&\|\bm W_{\rm t}^\ast- \bm F_{\rm t}\tilde{\bm W}_{\rm t}\|_F^2\\
    &\operatorname{subject~to}~&&\|\bm F_{\rm t}\tilde{\bm W}_{\rm t}\|_F^2\le N_{\rm s},\\ &&& |[\bm F_{\rm t}]_{ij}|=1, \forall i, j.
  \end{aligned}
\end{equation}
Similarly, the matrix decomposition problem for finding the hybrid beamformers at the receiver can be formulated as: 
\begin{equation}\label{eq:decomp2}
  \begin{aligned}
    &\underset{\bm F_{\rm r}, \tilde{\bm W}_{\rm r}}{\operatorname{minimize}}~&&\|\bm W_{\rm r}^\ast- \bm F_{\rm r}\tilde{\bm W}_{\rm r}\|_F^2\\
    &\operatorname{subject~to}~&& |[\bm F_{\rm r}]_{ij}|=1,\quad\forall i, j.
  \end{aligned}
\end{equation}
Both problem \eqref{eq:decomp1} and problem \eqref{eq:decomp2} are nonconvex due to the unit modulus constraints. Some heuristic algorithms are proposed to tackle the nonconvex hybrid beamforming problem \cite{7397861,7389996}, but in general finding the optimal solution can have high computational complexity. Furthermore, it is shown in \cite{7389996} that the perfect decomposition (with the optimal objective value $0$) exists when the number of RF chains is greater than twice the number of data streams, but it is not clear whether perfect decomposition exists when this sufficient condition is not met. Overall, the algorithm design for the hybrid structure is more difficult than the fully digital counterpart even when the CSI is available.

The hybrid antenna structure also makes channel sensing more challenging.
First, the limited number of RF chains in the hybrid receiver structure does
not allow a full observation of the high-dimensional channel output. This can
lead to a significant increase in pilot training overhead as compared to the
fully digital scenario. Second, the pilots are sent and received via the hybrid
architecture, which makes designing the sensing matrix more challenging due to
the additional nonconvex constraints in the hybrid structure.

To address the aforementioned challenges, we propose a new active sensing framework that is suitable for the hybrid beamforming scenario. The framework also depends on the assumption of channel reciprocity and the ping-pong pilot transmission protocol.

\subsection{Ping-Pong Pilots Transmission}
\begin{figure*}[t]
\centering
  \includegraphics[width=16cm]{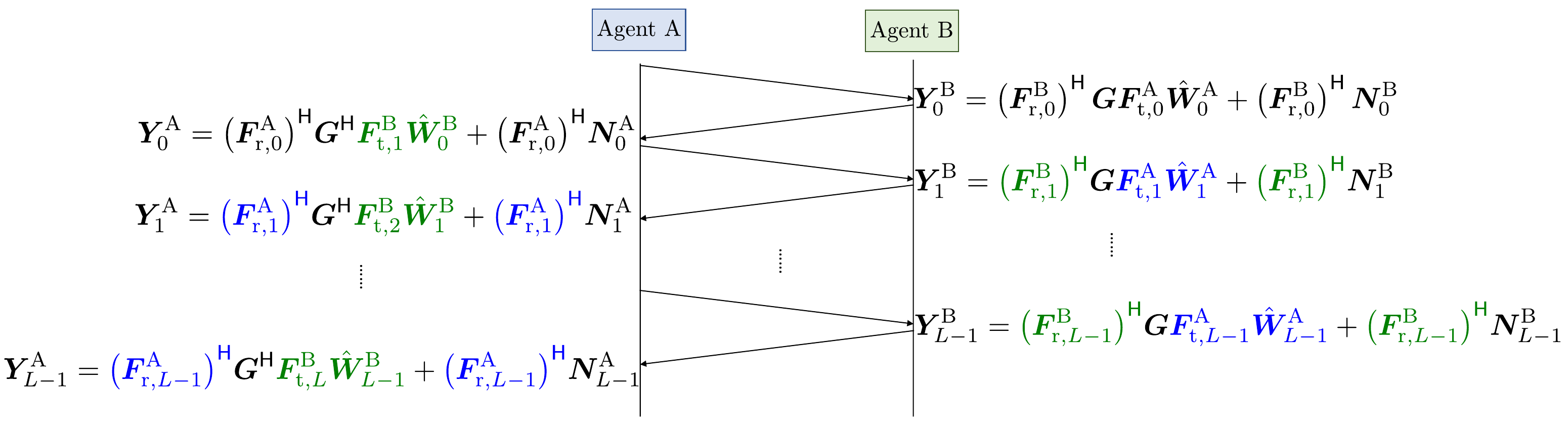}
  \definecolor{rx}{rgb}{0.0, 0.5, 0.0}
  \caption{Ping-pong pilot transmission protocol of $L$ rounds for hybrid MIMO. The sensing matrices adaptively designed at agent A and agent B are highlighted as green (e.g., \textcolor{rx}{$\hat{\bm W}_{0}^{\rm B}$}) and blue (e.g., \textcolor{blue}{$\hat{\bm W}_{1}^{\rm A}$}), respectively.}
  \label{fig:pingpong_pilots_hybrid}
\end{figure*}
The ping-pong pilot transmission protocol follows the one in Section \ref{sec:pingpongpilots} with the main difference in that the pilots are 
transmitted and received through a hybrid analog and digital structure.

Specifically, in the $\ell$-th ping-pong round, agent A sends a sequence of pilots of length $N_{\rm s}$, i.e., $\bm W_{\ell}^{\rm A}\in\mathbb{C}^{M_{\rm t}\times N_{\rm s}}$, which satisfies the hybrid beamforming constraint, i.e., $\bm W_{\ell}^{\rm A}=\bm F_{\rm t,\ell}^{\rm A}\hat{\bm W}_{\ell}^{\rm A}$, where $\bm F_{\rm t,\ell}^{\rm A}\in\mathbb{C}^{M_{\rm t}\times N_{\rm t}^{\rm RF}}$ is the analog phase shift\footnote{\textcolor{black}{Here, the analog phase shift matrices $\bm F_{\rm t,\ell}^{\rm A}, \bm F_{\rm r,\ell}^{\rm B}$ are kept the same for all the $N_{\rm s}$ pilots the in each round, i.e., for all the columns of \(\hat{\bm W}_{\ell}^{\rm A}\), to reduce the design and implementation complexity. The same applies to the transmission of pilots from agent B to agent A. It may be possible to enhance performance by designing different analog beamformers for different pilots.}} with $|[\bm F_{\rm t,\ell}^{\rm A}]_{ij}|=1$, $\forall~i,j$, and $\hat{\bm W}_{\ell}^{\rm A}\in\mathbb{C}^{N_{\rm t}^{\rm RF}\times N_{\rm s}}$ is the digital beamformed sequence. Each column in the overall beamforming matrix $\bm W_{\ell}^{\rm A}$ can be viewed as a pilot transmitted in one channel use and should satisfy a normalized power constraint. That is, the $i$-th column $\bm w_{\ell,i}^{\rm A}$ satisfies the power constraint $\|\bm w_{\ell,i}^{\rm A}\|_2\le 1$. At the receiver, since agent B has $N_{\rm r}^{\rm RF}$ RF chains, it can only observe $N_{\rm r}^{\rm RF}$ complex scalars which correspond to the signal after applying the analog combining matrix $\bm F_{\rm r,\ell}^{\rm A}$ in each channel use. Thus, the received pilots at agent B are given by:
\begin{equation}\label{eq:pilot_BA_hybrid}
  \bm Y_{\ell}^{\rm B} = \left(\bm F_{\rm r,\ell}^{\rm B}\right)^{\sf H}\bm G\bm F_{\rm t,\ell}^{\rm A}\hat{\bm W}_{\ell}^{\rm A}+ \left(\bm F_{\rm r,\ell}^{\rm B}\right)^{\sf H}\bm N_{\ell}^{\rm B},\quad \ell=0,\ldots,L-1,
\end{equation}
where $\bm N_{\ell}^{\rm B}$ is the additive white Gaussian noise with each column independently distributed as $\mathcal{CN}(\bm 0,\sigma^{2}_{\rm B}\bm I)$. Here, the analog combining matrix has the constraint $|[\bm F_{\rm r,\ell}^{\rm B}]_{ij}|=1$, $\forall i,j$.

Given the received pilots $\bm Y_{\ell}^{\rm B}$, agent B exploits all the historical observations to update the next sensing beamformers $\hat{\bm W}_{\ell}^{\rm B},\bm F_{\rm t,\ell+1}^{\rm B},\bm F_{\rm r,\ell+1}^{\rm B}$ as follows:
\begin{subequations}\label{eq:sensing_B_hybrid}
  \begin{align}
    &\hat{\bm W}_{\ell}^{\rm B}= f_{{\rm d}, \ell}^{\rm B}([\bm Y_{0}^{\rm B},\ldots,\bm Y_{\ell}^{\rm B}]),\quad\ell=0,\ldots,L-1,\\
    &\bm F_{\rm t,\ell+1}^{\rm B}= f_{{\rm t}, \ell}^{\rm B}([\bm Y_{0}^{\rm B},\ldots,\bm Y_{\ell}^{\rm B}]),\quad\ell=0,\ldots,L-1,\\
    &\bm F_{\rm r,\ell+1}^{\rm B}= f_{{\rm r}, \ell}^{\rm B}([\bm Y_{0}^{\rm B},\ldots,\bm Y_{\ell}^{\rm B}]),\quad\ell=0,\ldots,L-2,
  \end{align}
\end{subequations}
where $ f_{{\rm d}, \ell}^{\rm B}:\mathbb{C}^{ N_{\rm r}^{\rm RF} \times (\ell+1)N_{\rm s}}\rightarrow \mathbb{C}^{N_{\rm r}^{\rm RF}\times N_{\rm s}} $ maps the received pilots to the next digital transmit beamformer, $f_{{\rm t}, \ell}^{\rm B}:\mathbb{C}^{N_{\rm r}^{\rm RF}\times (\ell+1)N_{\rm s}}\rightarrow \mathbb{C}^{M_{\rm r}\times N_{\rm r}^{\rm RF}} $
maps the received pilots to the next analog transmit beamformer, and $f_{{\rm r}, \ell}^{\rm B}:\mathbb{C}^{N_{\rm r}^{\rm RF}\times (\ell+1)N_{\rm s}}\rightarrow \mathbb{C}^{M_{\rm r}\times N_{\rm r}^{\rm RF}} $ maps the received pilots to the next analog receive beamformer.  

Next, the agent B constructs the pilot matrix as $\bm W_{\ell}^{\rm B}=\bm F_{{\rm t},\ell+1}^{\rm B}\hat{\bm W}_{\ell}^{\rm B}$ where each column of the matrix $\bm W_{\ell}^{\rm B}$ has a transmit power constraint, i.e., $\|\bm w_{\ell,i}^{\rm B}\|_2\le 1$, and sends the pilot to agent A. 
The pilot sequence received at agent A is given by
\begin{equation}\label{eq:pilot_AB_hybrid}
  \bm Y_{\ell}^{\rm A} = \left(\bm F_{\rm r,\ell}^{\rm A}\right)^{\sf H}\bm G^{\sf H}\bm F_{\rm t,\ell+1}^{\rm B}\hat{\bm W}_{\ell}^{\rm B}+ \left(\bm F_{\rm r, \ell}^{\rm A}\right)^{\sf H}\bm N_{\ell}^{\rm A},\quad \ell=0,\ldots,L-1,
\end{equation}
where $\bm N_{\ell}^{\rm A}$ is the additive white Gaussian noise with each column independently distributed as $\mathcal{CN}(\bm 0,\sigma^{2}_{\rm A}\bm I)$, and $\bm F_{\rm r, \ell}^{\rm A}$ is the analog receive beamformer at agent A, and satisfies the constraint $|[\bm F_{\ell}^{\rm A}]_{ij}|=1$, $\forall~i,j$. 
Upon receiving the pilots $\bm Y_{\ell}^{\rm A}$, agent A designs its next analog and digital beamformers as follows:
\begin{subequations}\label{eq:sensing_A_hybrid}
  \begin{align}
    &\hat{\bm W}_{\ell+1}^{\rm A}= f_{{\rm d}, \ell}^{\rm A}([\bm Y_{0}^{\rm A},\ldots,\bm Y_{\ell}^{\rm A}]),\quad\ell=0,\ldots,L-2,\\
    &\bm F_{\rm t, \ell+1}^{\rm A}= f_{{\rm t}, \ell}^{\rm A}([\bm Y_{0}^{\rm A},\ldots,\bm Y_{\ell}^{\rm A}]),\quad\ell=0,\ldots,L-2,\\
    &\bm F_{\rm r,\ell+1}^{\rm A}= f_{{\rm r}, \ell}^{\rm A}([\bm Y_{0}^{\rm A},\ldots,\bm Y_{\ell}^{\rm A}]),\quad\ell=0,\ldots,L-2,
  \end{align}
\end{subequations}
where $ f_{{\rm d}, \ell}^{\rm A}:\mathbb{C}^{ N_{\rm t}^{\rm RF} \times (\ell+1)N_{\rm s}}\rightarrow \mathbb{C}^{N_{\rm t}^{\rm RF}\times N_{\rm s}}  $, $f_{{\rm t}, \ell}^{\rm A}:\mathbb{C}^{ N_{\rm t}^{\rm RF} \times (\ell+1)N_{\rm s}}\rightarrow \mathbb{C}^{M_{\rm t}\times N_{\rm t}^{\rm RF}} $ and $f_{{\rm r}, \ell}^{\rm A}:\mathbb{C}^{ N_{\rm t}^{\rm RF} \times (\ell+1)N_{\rm s}}\rightarrow \mathbb{C}^{M_{\rm t}\times N_{\rm t}^{\rm RF}} $ map the received pilots to the corresponding sensing beamformers in the $(\ell+1)$-th round, respectively. 

The overall ping-pong pilots transmission protocol is illustrated in Fig.~\ref{fig:pingpong_pilots_hybrid}. After the pilot sensing stage, the final data transmission beamformers are designed as follows:
\begin{subequations}\label{eq:data_hybrid}
  \begin{align}
    &\hat{\bm W}_{\rm t}= g_{\rm d}^{\rm A}([\bm Y_{0}^{\rm A},\ldots,\bm Y_{L-1}^{\rm A}]),\\
    &\hat{\bm W}_{\rm r}= g_{\rm d}^{\rm B}([\bm Y_{0}^{\rm B},\ldots,\bm Y_{L-1}^{\rm B}]),\\
    &\bm F_{\rm t} =  g_{\rm a}^{\rm A}([\bm Y_{0}^{\rm A},\ldots,\bm Y_{L-1}^{\rm A}]),\\
    &\bm F_{\rm r} =  g_{\rm a}^{\rm B}([\bm Y_{0}^{\rm B},\ldots,\bm Y_{L-1}^{\rm B}]),
  \end{align}
\end{subequations}
where $g_{\rm d}^{\rm A}, g_{\rm d}^{\rm B}, g_{\rm a}^{\rm A}, g_{\rm a}^{\rm B}$ are functions that map all the received pilots to the data transmission and receiving beamformers, the analog beamformers $\bm F_{\rm t}$ and  $\bm F_{\rm r}$ have unit modulus constraints, and the overall transmit beamformer satisfies a power constraint $\|\bm F_{\rm t}\tilde{\bm W}_{\rm t}\|_F^2\le N_{\rm s}$. The overall active sensing problem can be formulated as: 
\begin{equation}\label{eq:formulation2_hybrid}
  \begin{aligned}
    &\underset{\mathcal{F}}{\operatorname{maximize}}~&&\mathbb{E}\left[\log|\det(\bm F_{\rm r}^{\sf H}\hat{\bm W}_{\rm r}^{\sf H} \bm G\bm F_{\rm t}\hat{\bm W}_{\rm t})|^2\right]\\
    &\operatorname{subject~to}~&& \eqref{eq:sensing_B_hybrid},\eqref{eq:sensing_A_hybrid}, \eqref{eq:data_hybrid},
  \end{aligned}
\end{equation}
where $\mathcal{F}=\{\{f_{\rm d,\ell}^{\rm A}\}_{\ell=0}^{L-2}, \{f_{\rm t,\ell}^{\rm A}\}_{\ell=0}^{L-2}, \{f_{\rm r,\ell}^{\rm A}\}_{\ell=0}^{L-2}, \{f_{\rm d,\ell}^{\rm B}\}_{\ell=0}^{L-1},\allowbreak  \{f_{\rm t, \ell}^{\rm B}\}_{\ell=0}^{L-1},  \{f_{\rm r, \ell}^{\rm B}\}_{\ell=0}^{L-2}, g_{\rm d}^{\rm A},g_{\rm d}^{\rm B},g_{\rm a}^{\rm A},g_{\rm a}^{\rm B}\}$ consists of all the mappings for designing the sensing strategy. Compared to the fully digital MIMO scenario in \eqref{eq:formulation2}, problem \eqref{eq:formulation2_hybrid} is more challenging due to the additional functional mappings and the nonconvex constraints.

\section{Proposed Active Sensing Framework for Hybrid MIMO System}\label{sec:method_hybrid}
\begin{figure*}[t]
  \centering
  \includegraphics[width=15cm]{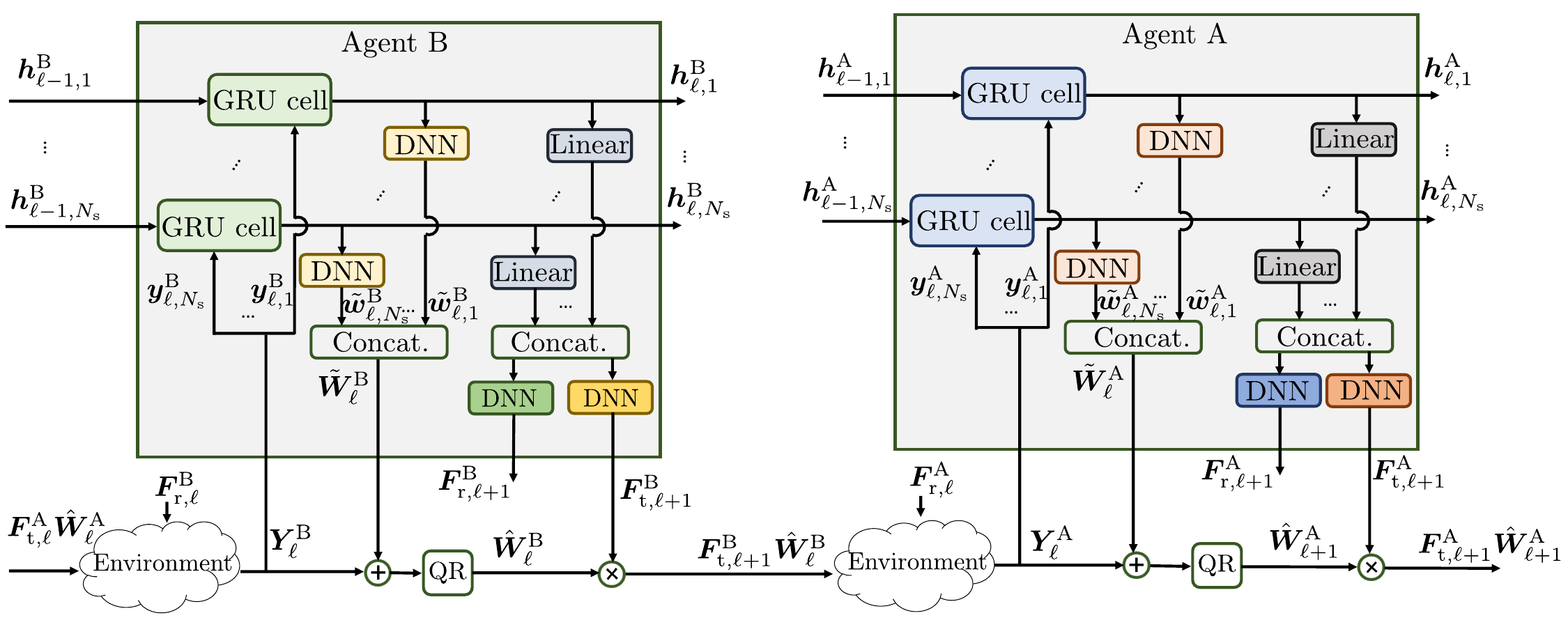}
  \caption{Proposed active sensing unit in the $\ell$-th pilot round for hybrid MIMO.}
  \label{fig:GRU_unit_hybrid}
\end{figure*}
\begin{algorithm}[t]
  \begin{algorithmic}[1]
      \STATE Initial $\bm W_{0}^{\rm A}\in\mathbb{C}^{M_{t}\times N_{\rm s}}$, $\bm h_{-1,i}^{\rm B}\in\mathbb{C}^{N_{\rm h}^{\rm B}}$, $\bm h_{-1,i}^{\rm A}\in\mathbb{C}^{N_{\rm h}^{\rm A}}$.
      \FOR{$\ell=0,\ldots,L-1$} 
      \STATE{A$\rightarrow$ B:\quad $\bm Y_{\ell}^{\rm B} = (\bm F_{\rm r,\ell}^{\rm B})^{\sf H}\bm G\bm F_{\rm t,\ell}^{\rm A}\tilde{\bm W}_{\ell}^{\rm A}+ (\bm F_{\rm r,\ell}^{\rm B})^{\sf H}\bm N_{\ell}^{\rm B}$} 
      \FOR{$i=1,\ldots,N_{\rm s}$}
      \STATE{$\bm h_{\ell,i}^{\rm B}=\operatorname{GRUCell}^{\rm B}(\bm h_{\ell-1,i}^{\rm B},\bm y_{\ell,i}^{\rm B})$}
      \STATE{$\tilde{\bm w}_{\ell,i}^{\rm B} = \operatorname{DNN}^{\rm B}(\bm h_{\ell,i}^{\rm B})$}
      \STATE{$\bm z_{\ell,i}^{\rm B} = \operatorname{Linear}(\bm h_{\ell,i}^{\rm B})$}
      \ENDFOR
      \STATE{$\tilde{\bm W}_{\ell}^{\rm B}=[\tilde{\bm w}_{\ell,1}^{\rm B},\ldots,\tilde{\bm w}_{\ell,N_{\rm s}}^{\rm B}]$}
      \STATE QR decomposition: {($\tilde{\bm W}_{\ell}^{\rm B}+\bm Y_{\ell}^{\rm B}) = \bm Q^{\rm B}_{\ell}\bm R_{\ell}^{\rm B}$}
      \STATE{Set $\hat{\bm W}_{\ell}^{\rm B} = \bm Q^{\rm B}_{\ell}.$}

      \STATE{$\bm F_{\rm r,\ell+1}^{\rm B} = \operatorname{DNN}_{\rm r}^{\rm B}([(\bm z_{\ell,1}^{\rm B})^\top,\cdots,(\bm z_{\ell,N_{\rm s}}^{\rm B})^\top]^\top)$}
      \STATE{$\bm F_{\rm t,\ell+1}^{\rm B} = \operatorname{DNN}_{\rm t}^{\rm B}([(\bm z_{\ell,1}^{\rm B})^\top,\cdots,(\bm z_{\ell,N_{\rm s}}^{\rm B})^\top]^\top)$}
            
      \STATE{}
      \STATE{B$\rightarrow$ A:\quad $\bm Y_{\ell}^{\rm A} = (\bm F_{\rm r,\ell}^{\rm A})^{\sf H}\bm G^{\sf H}\bm F_{\rm t,\ell+1}^{\rm B}\hat{\bm W}_{\ell}^{\rm B}+ (\bm F_{\rm r, \ell}^{\rm A})^{\sf H}\bm N_{\ell}^{\rm A}$} 
      \FOR{$i=1,\ldots,N_{\rm s}$}
      \STATE{$\bm h_{\ell,i}^{\rm A}=\operatorname{GRUCell}^{\rm A}(\bm h_{\ell-1,i}^{\rm A},\bm y_{\ell,i}^{\rm A})$}
      \STATE{$\tilde{\bm w}_{\ell,i}^{\rm A} = \operatorname{DNN}^{\rm A}(\bm h_{\ell,i}^{\rm A})$}
      \STATE{$\bm z_{\ell,i}^{\rm A} = \operatorname{Linear}(\bm h_{\ell,i}^{\rm A})$}
      \ENDFOR
      \STATE{$\tilde{\bm W}_{\ell}^{\rm A}=[\tilde{\bm w}_{\ell,1}^{\rm A},\ldots,\tilde{\bm w}_{\ell,N_{\rm s}}^{\rm A}]$}
      \STATE QR decomposition: {($\tilde{\bm W}_{\ell}^{\rm A}+\bm Y_{\ell}^{\rm A}) = \bm Q^{\rm A}_{\ell}\bm R_{\ell}^{\rm A}$}
      \STATE{Set $\hat{\bm W}_{\ell+1}^{\rm A} = \bm Q^{\rm A}_{\ell}$}
      \STATE{$\bm F_{\rm r,\ell+1}^{\rm A} = \operatorname{DNN}_{\rm r}^{\rm A}([(\bm z_{\ell,1}^{\rm A})^\top,\cdots,(\bm z_{\ell,N_{\rm s}}^{\rm A})^\top]^\top)$}
      \STATE{$\bm F_{\rm t,\ell+1}^{\rm A} = \operatorname{DNN}_{\rm t}^{\rm A}([(\bm z_{\ell,1}^{\rm A})^\top,\cdots,(\bm z_{\ell,N_{\rm s}}^{\rm A})^\top]^\top)$}
      \ENDFOR
  \end{algorithmic}
  \caption{Proposed active sensing framework for hybrid MIMO.}\label{algo2}
\end{algorithm}
This section proposes an active sensing framework to adaptively design the analog and digital beamforming matrices for the hybrid MIMO system. The general idea is to utilize the proposed active sensing framework in Section \ref{sec:active_unit} to design the digital sensing beamformers for the effective channel with fixed analog beamformers, while actively updating the analog beamformers in each ping-pong round by utilizing the learned hidden state vectors.

The detail of the active sensing unit for the hybrid scenario is illustrated in Fig.~\ref{fig:GRU_unit_hybrid}. The active sensing unit consists of $N_{\rm s}$ GRU cells at each side to abstract information from the received pilots into their hidden states, and a set of DNNs to map the latest hidden states to the next analog and digital beamformers. In particular, each GRU at agent B takes one  column of the received pilots $ \bm Y_{\ell}^{\rm B} $ as input to update its hidden states as follows:
\begin{equation}\label{eq:update_hidden_hybrid}
\bm h_{\ell,i}^{\rm B}=\operatorname{GRUCell}^{\rm B}(\bm h_{\ell-1,i}^{\rm B},\bm y_{\ell,i}^{\rm B}),\quad i=1,\ldots,N_{\rm s},
\end{equation}
where $\operatorname{GRUCell}^{\rm B}$ follows the same implementation in \eqref{eq:gru_imp} but the input dimension is adjusted to accommodate the dimension of $\bm y_{\ell,i}^{\rm B}\in\mathbb{C}^{N_{\rm r}^{\rm RF}}$ in the hybrid scenario.  

To design the next  digital sensing beamformer $\hat{\bm W}_{\ell}^{\rm B}$, we follow the similar steps in \eqref{eq:tilde_w},\eqref{eq:cat_w},\eqref{eq:final_W} as follows:
\begin{subequations}
\begin{align}
&\tilde{\bm w}_{\ell,i}^{\rm B} = \operatorname{DNN}^{\rm B}(\bm h_{\ell,i}^{\rm B}),\quad i=1,\ldots,N_{\rm s}. \label{eq:tilde_w_hybrid}\\
&\tilde{\bm W}_{\ell}^{\rm B}=[\tilde{\bm w}_{\ell,1}^{\rm B},\ldots,\tilde{\bm w}_{\ell,N_{\rm s}}^{\rm B}],\\
    &(\tilde{\bm W}_{\ell}^{\rm B}+\bm Y_{\ell}^{\rm B}) = \bm Q^{\rm B}_{\ell}\bm R_{\ell}^{\rm B} \quad(\text{QR decomposition}),\\
    &\hat{\bm W}_{\ell}^{\rm B} = \bm Q^{\rm B}_{\ell}.
\end{align}
\end{subequations}
We note that the output dimension of $\operatorname{DNN}^{\rm B}$ needs to be adjusted to accommodate the dimension of $\tilde{\bm w}_{\ell,i}^{\rm B}\in\mathbb{C}^{N_{\rm r}^{\rm RF}}$. 

To design the next analog transmit beamformer $\bm F_{\rm t,\ell+1}^{\rm B}$ and analog receive beamformer $\bm F_{\rm r,\ell+1}^{\rm B}$, we first apply a linear layer to each hidden state vector to reduce its dimension to $N_{\rm f}^{\rm B}$, then concatenate the output of the linear layer to form a feature vector, which is taken as an input to a DNN for producing the analog beamformer as follows:
\begin{subequations}\label{eq:final_W_hybrid}
  \begin{align}
    &\bm z_{\ell,i}^{\rm B} = \operatorname{Linear}(\bm h_{\ell,i}^{\rm B}),~i=1,\cdots,N_{\rm s},\\
    &\bm F_{\rm r,\ell+1}^{\rm B} = \operatorname{DNN}_{\rm r}^{\rm B}([(\bm z_{\ell,1}^{\rm B})^\top,\cdots,(\bm z_{\ell,N_{\rm s}}^{\rm B})^\top]^\top),\\
    &\bm F_{\rm t, \ell+1}^{\rm B} = \operatorname{DNN}_{\rm t}^{\rm B}([(\bm z_{\ell,1}^{\rm B})^\top,\cdots,(\bm z_{\ell,N_{\rm s}}^{\rm B})^\top]^\top),
  \end{align}
\end{subequations}
We note that the analog beamformers are designed based on the concatenation of information from all the GRU hidden states because the common analog beamformers are shared by the digital beamformers corresponding to different data streams.

The hybrid sensing beamformers at agent A are similarly designed. The overall framework at the sensing stage is summarized in Algorithm~\ref{algo2}. After the channel sensing stage, agent A and agent B exploit the information obtained at the sensing stage to design the hybrid beamformers for data transmission and reception, respectively. Specifically, after $L$-rounds pilot transmission, the hybrid receive beamformer at agent B is given by:
\begin{subequations}\label{eq:data_hybrid_dnn}
  \begin{align}
    &\tilde{\bm w}_{{\rm r},i} =\operatorname{DNN}^{\rm r}(\bm h_{L,i}^{\rm B}),\quad i=1,\ldots,N_{\rm s},\label{eq:dnn_r_hybrid}\\
    &\tilde{\bm W}_{{\rm r}} = [\tilde{\bm w}_{{\rm r},1},\ldots,\tilde{\bm w}_{{\rm r},N_{\rm s}}],\\
    &(\tilde{\bm W}_{{\rm r}}+\bm Y_{L-1}^{\rm B}) = \bm Q^{\rm r}\bm R^{\rm r},\quad(\text{QR decomposition}),\\
    &\hat{\bm W}_{{\rm r}} = \bm Q^{\rm r},\\
    &\bm F_{\rm r} =  \bm F_{\rm r, L}^{\rm B},
  \end{align}
\end{subequations}
We note that a fully connected neural network $\operatorname{DNN}^{\rm r}$ is used to design the digital beamformer $\hat{\bm W}_{\rm r}$. In contrast, the analog beamformer $\bm F_{\rm r}$ is set as the latest analog receive beamformer $\bm F_{\rm r, L}^{\rm B}$ without introducing an additional DNN. This design reduces training complexity while achieving satisfactory performance, as observed in simulations.
Agent A follows the same steps in \eqref{eq:data_hybrid_dnn} to design its analog beamformer $\bm F_{\rm t}$ and digital beamformer $\hat{\bm W}_{\rm t}$. Each column of the matrices $\bm F_{\rm t}\hat{\bm W}_{\rm t}$ and $\bm F_{\rm r}\hat{\bm W}_{\rm r}$ are normalized to unit 2-norm to meet the power constraint.
We concatenate $L$ active sensing units together to train the overall network end-to-end with the loss function
$-\mathbb{E}[\log|\det(\bm F_{\rm r}^{\sf H}\hat{\bm W}_{\rm r}^{\sf H} \bm G\bm F_{\rm t}\hat{\bm W}_{\rm t})|^2]$. For better generalization to the ping-pong rounds $L$, the loss function can be set as $-\mathbb{E}[\sum_{\ell=0}^{L-1}\log|\det(\bm F_{\rm r,\ell}^{\sf H}\hat{\bm W}_{\rm r,\ell}^{\sf H} \bm G\bm F_{\rm t,\ell}\hat{\bm W}_{\rm t,\ell})|^2]$, where $\bm F_{\rm r,\ell},\hat{\bm W}_{\rm r,\ell},\bm F_{\rm t,\ell}$ and $\hat{\bm W}_{\rm t,\ell}$ are the data reception and transmission beamformers generated from the hidden states of the $\ell$-th round.

\section{Performance Evaluation on Hybrid MIMO}\label{sec:simulation_hybrid}
\begin{figure*}
     \centering
     \subfloat[SNR=$-5$dB, $N_{\rm s}=2$]{\includegraphics[width=0.32\textwidth]{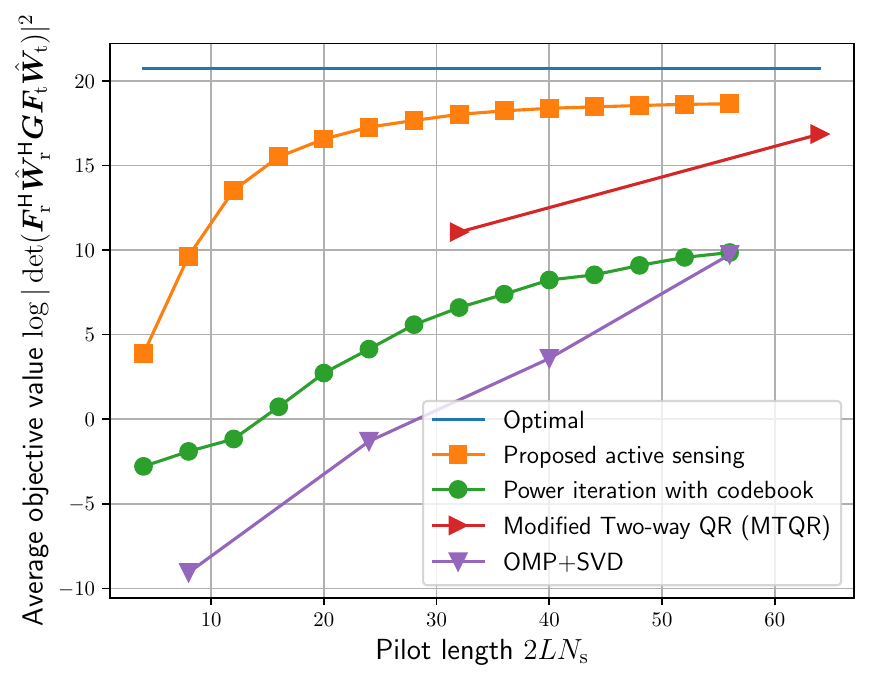}\label{fig:hybrid_snr-5_2}}
\hfil
\subfloat[SNR=$0$dB, $N_{\rm s}=2$]{\includegraphics[width=0.32\textwidth]{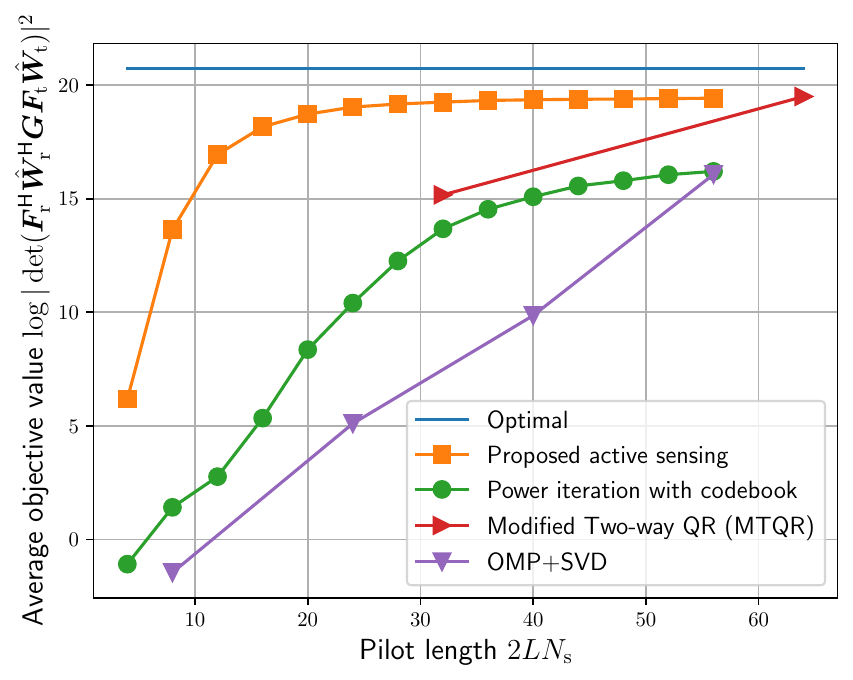}\label{fig:hybrid_snr0_2}}
\hfil
\subfloat[SNR=$5$dB, $N_{\rm s}=2$]{\includegraphics[width=0.32\textwidth]{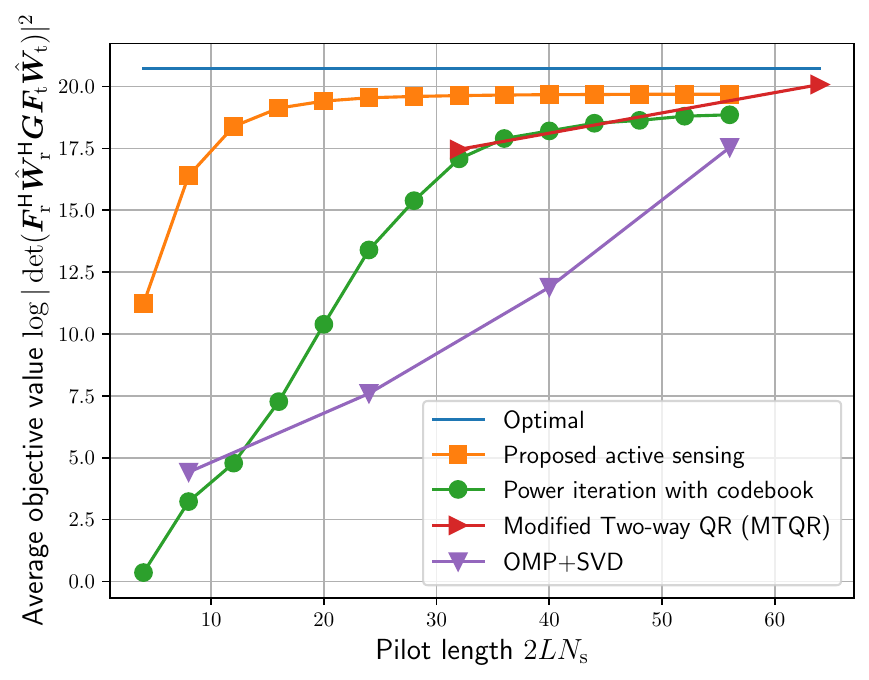}\label{fig:hybrid_snr5_2}}
\\ 
\subfloat[SNR=$0$dB, $N_{\rm s}=4$]{\includegraphics[width=0.32\textwidth]{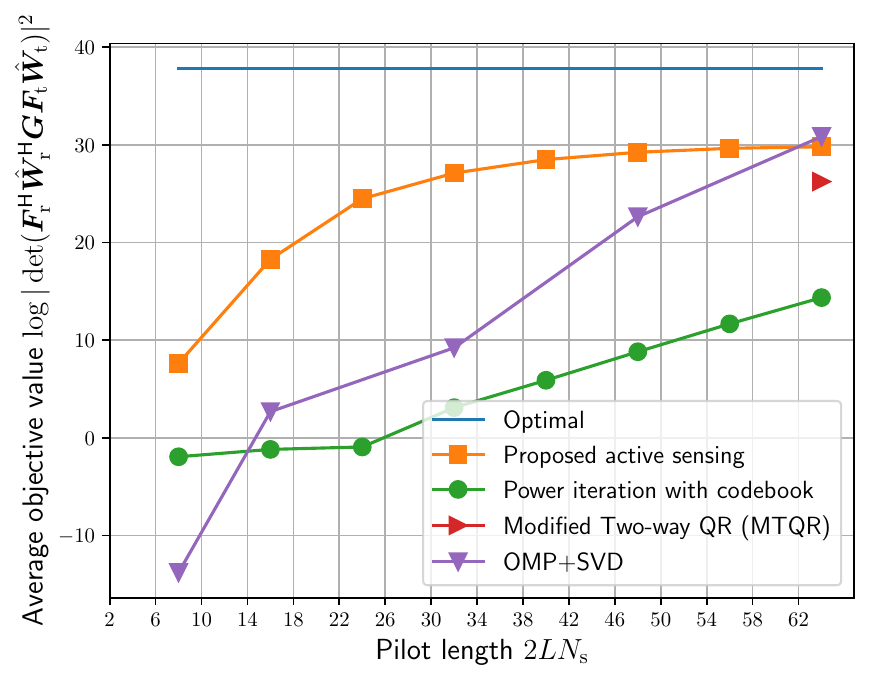}\label{fig:hybrid_snr0_4}}
\hfil
\subfloat[SNR=$5$dB, $N_{\rm s}=4$]{\includegraphics[width=0.32\textwidth]{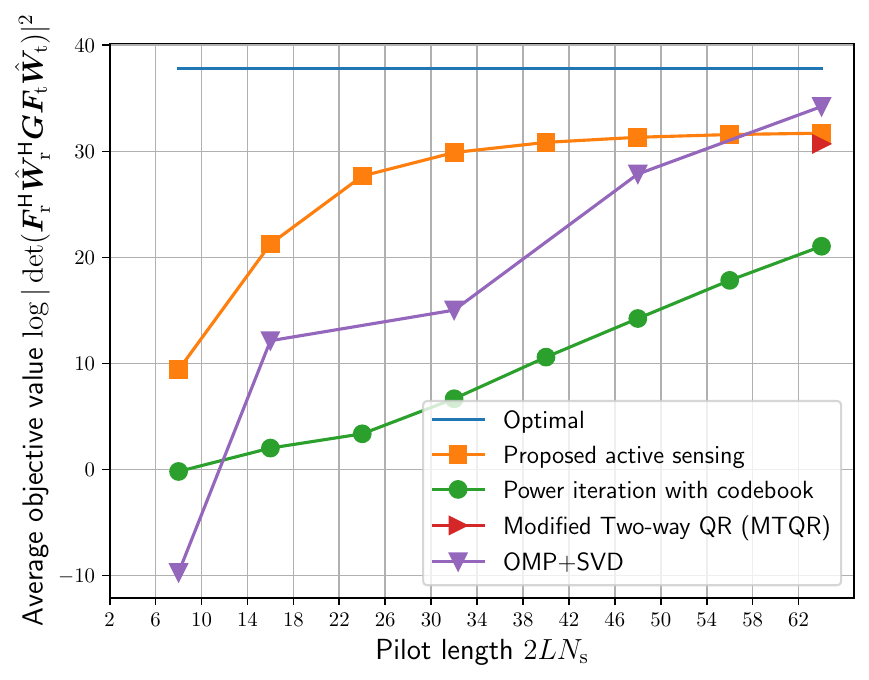}\label{fig:hybrid_snr5_4}}
\hfil
\subfloat[SNR=$10$dB, $N_{\rm s}=4$]{\includegraphics[width=0.32\textwidth]{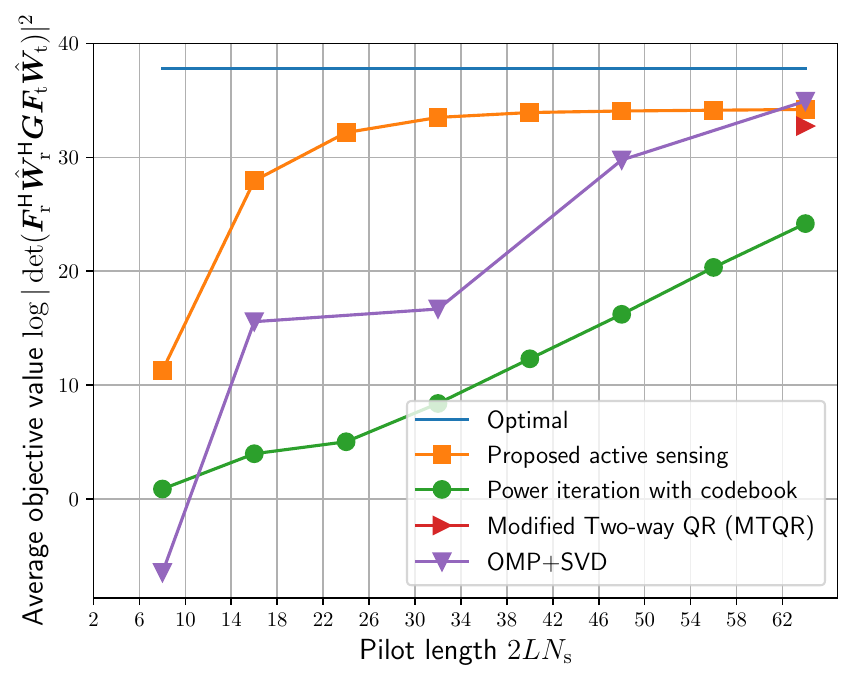}\label{fig:hybrid_snr10_4}}
     \caption{Performance comparison under a sparse mmWave channel model for a hybrid MIMO system with $M_{\rm t}=M_{\rm r}=64$ and $N_{\rm t}^{\rm RF}=N_{\rm r}^{\rm RF}=8$. }
     \label{fig:hybrid_sim}
\end{figure*}

In this section, we evaluate the performance of the proposed active sensing framework on a point-to-point MIMO system with a hybrid antenna structure. The number of antennas at agent A and agent B are $M_{\rm t}=64$ and $M_{\rm r}=64$, respectively. Both agents have $8$ RF chains, i.e., $N_{\rm t}^{\rm RF}=N_{\rm r}^{\rm RF}=8$. We consider a mmWave channel model as follows:
\begin{equation}
    \bm G = \sum_{\ell=1}^{L_{\rm p}}\alpha_\ell\bm a(\theta_\ell)\bm b(\phi_\ell)^{\top},
\end{equation}
where $\alpha_\ell\sim\mathcal{CN}(0,1)$ is the fading coefficient, 
$\theta_\ell\sim \text{Uniform}(-60^\circ,60^\circ)$ and $\phi_\ell\sim \text{Uniform}(-60^\circ,60^\circ)$ are angle-of-arrivals (AoAs) and angle-of-departures (AoDs), respectively, $\bm a(\theta_\ell)=[1,\ldots,e^{j\pi(M_{\rm r}-1)\sin(\theta_\ell)}]^\top$  and $\bm b(\phi_\ell)=[1,\ldots,e^{j\pi(M_{\rm t}-1)\sin(\phi_\ell)}]^\top$ are steering vectors assuming uniform linear array with half-wavelength antenna spacing. We set $L_{\rm p} = 4$ for the $N_{\rm s}=2$ scenario and $L_{\rm p} = 10$ for the $N_{\rm s}=4$ case.

To implement the active sensing framework in Algorithm~\ref{algo2}, the dimensions of the hidden state from both sides are set as $N_{\rm h}^{\rm A}=N_{\rm h}^{\rm B}=512$. All the fully connected neural networks have two hidden layers of dimension $512$ and with $\operatorname{relu}(\cdot)$ activation function. We set $N_{\rm f}^{A} = N_{\rm f}^B =128 $. The dimensions of the input and output layers are set according to the functionality of the DNN. The loss function is set as $-\mathbb{E}[\sum_{\ell=0}^{L-1}\log|\det(\bm F_{\rm r,\ell}^{\sf H}\hat{\bm W}_{\rm r,\ell}^{\sf H} \bm G\bm F_{\rm t,\ell}\hat{\bm W}_{\rm t,\ell})|^2]$. During training, the number of ping-pong rounds $L$ is set to be $8$. The neural network is trained in the same way as in section \ref{sec:performance_digital_mimo}. 

The proposed active sensing framework is compared with the following benchmarks:

\textit{SVD with Perfect CSI:} Given perfect CSI $\bm G$, the overall precoding matrix $\bm W_{\rm t}$ and combining matrix $\bm W_{\rm r}$ are given by the optimal solution in \eqref{eq:opt_precoding_decoding}. Here, in the simulation scenarios with $N_{\rm t}^{\rm RF}=N_{\rm r}^{\rm RF}=8$ and $N_{\rm s} = 2$ or $4$,
the optimal digital and analog beamformers can always be obtained from the decomposition of the overall precoding and combing matrices.

\textit{Power Iteration with Codebook\cite{8644444}:} The scheme uses the same ping-pong pilot protocol as in this paper. The digital beamformers are designed based on the power iteration method proposed in \cite{1323251}, and the analog beamformers are adaptively selected from a predefined codebook using a “multi-beam split and drop with backtracking” method.

\textit{OMP+SVD\cite{6847111}:} In this scheme, agent A sends $2LN_{\rm s}$ pilots to agent B with random transmit and receive sensing vectors. Agent B first adopts the orthogonal matching pursuit (OMP) algorithm \cite{4385788} to estimate the channel by exploiting the sparsity structure of the channel then performs SVD on the estimated channel to obtain the overall precoding matrix $\bm W_{\rm t}$ and combining matrix $\bm W_{\rm r}$ as in \eqref{eq:opt_precoding_decoding}. Agent B then sends the designed $\bm W_{\rm t}$ to agent A through a noiseless feedback channel. Here again, the optimal digital and analog beamformers can be obtained from the decomposition of the overall precoding and combining matrices.

\textcolor{black}{\textit{Modified Two-Way QR (MTQR) \cite{7439748}:}  The MTQR algorithm employs a repetition transmission scheme, utilizing the same sensing beamformers \(N_{\rm t}/N_{\rm RF}\) or \(N_{\rm r}/N_{\rm RF}\) times to accumulate a total of \(N_{\rm t}\) and \(N_{\rm r}\) observations at agent A and B, respectively. This is followed by the power iteration method to design the next set of overall hybrid sensing beamformers, which are then decomposed into separate digital and analog beamformers at agent A and agent B, respectively.}

In Fig.~\ref{fig:hybrid_sim}, we present the simulation results for different SNRs with the number of data streams $N_{\rm s}=2$ or $N_{\rm s}=4$. We observe that the proposed active sensing method achieves the best performance in almost all the scenarios especially when the SNR is low and the number of data streams is small. For example, it can be seen from Fig.~\ref{fig:hybrid_snr-5_2} that the proposed active sensing method significantly outperforms the other methods, while power iteration with codebook only appears relatively advantageous over the channel estimation based scheme. As the SNR increases (e.g., in Fig.~\ref{fig:hybrid_snr0_2} and \ref{fig:hybrid_snr5_2}) the performance of power iteration with codebook gradually approaches the proposed active sensing scheme, but the proposed scheme still converges much faster as the number of pilots transmission increases, thereby achieving lower pilot overhead. For instance, the proposed scheme with $16$ pilots obtains better performance than the other benchmarks with $56$ pilots. \textcolor{black}{ Additionally, we observe that the MTQR algorithm can achieve relatively good performance with a sufficiently large number of pilots, however, it is not feasible when the number of accumulated pilot samples is less than the number of receiving antennas.}

We also present the simulation results for the scenario with $N_{\rm s}=4$ data streams in 
Fig.~\ref{fig:hybrid_snr0_4}--\ref{fig:hybrid_snr10_4}. It can be observed that the performance gain of active sensing is smaller than the case of $N_{\rm s}=2$. The power iteration method with codebook performs even worse than the channel estimation based scheme, which randomly senses the channel. However, we still observe
considerable performance gain of the proposed active sensing method over the channel estimation based approach especially when the pilot length is small. We also observe that the channel estimation based approach performs slightly better than the active sensing scheme when the pilot length is $64$. This is because if the pilot length is sufficiently large the two agents already have enough observations about the channel from the random sensing stage. This implies that the active sensing method is most suitable for the scenario when the pilot length is short. We also remark that the OMP-based channel estimation assumes the AoA and AoD come from a $64\times 64$ grid in simulations, so its actual performance would degrade if evaluated on the actual gridless channel, while other methods do not make such an assumption. 

\section{Discussions}
\subsection{Interpretation of the Learned Solutions}\label{sec:interpretation}
\begin{figure*}[t]
  \centering
  \subfloat[Proposed, SNR=$0$dB]{\includegraphics[width=0.24\textwidth]{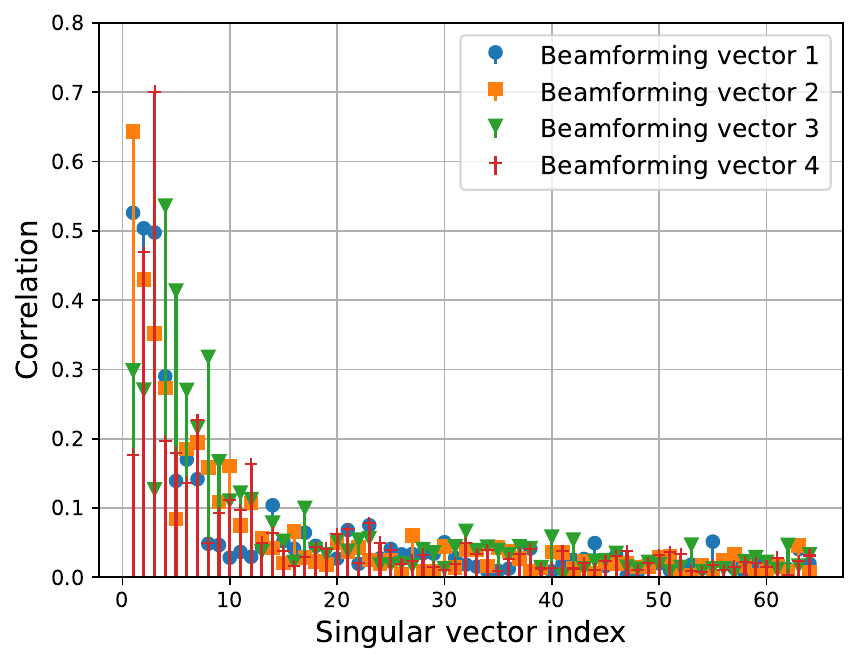}\label{fig:inter_d_a}}
  \hfil
  \subfloat[Power iteration, SNR=$0$dB]{\includegraphics[width=0.24\textwidth]{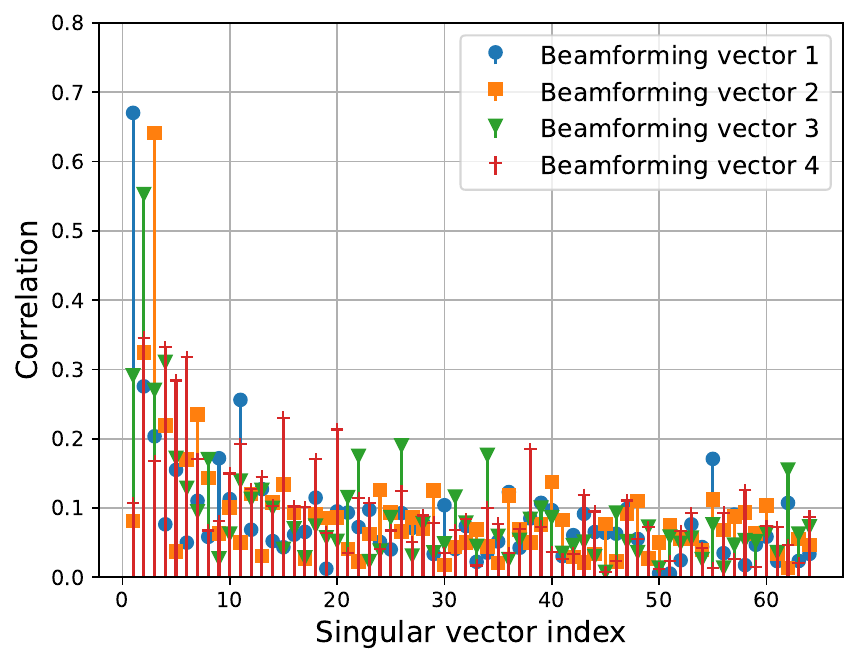}\label{fig:inter_d_b}}
  \hfil
  \subfloat[Proposed, SNR=$-5$dB]{\includegraphics[width=0.24\textwidth]{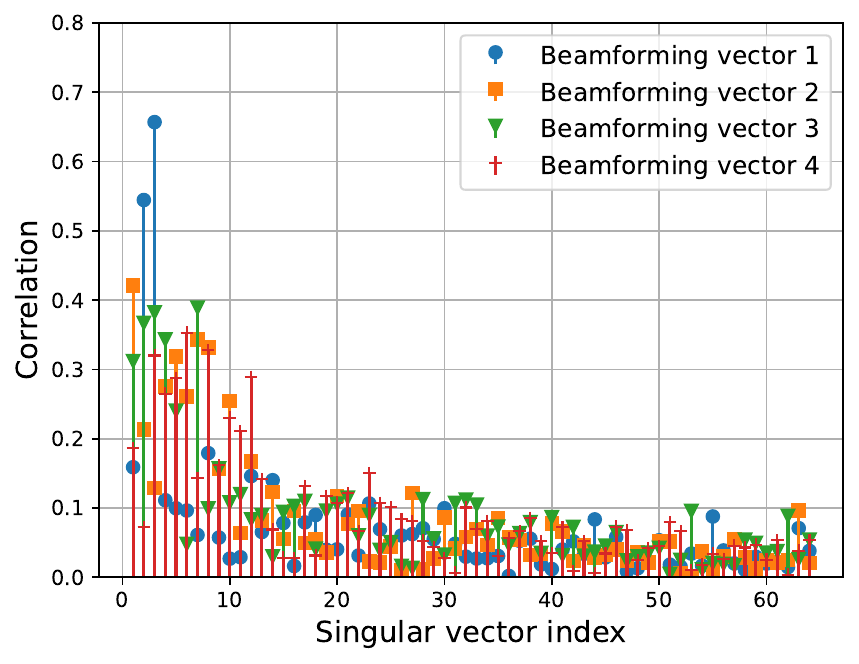}\label{fig:inter_d_c}}
  \hfil
  \subfloat[Power iteration, SNR=$-5$dB]{\includegraphics[width=0.24\textwidth]{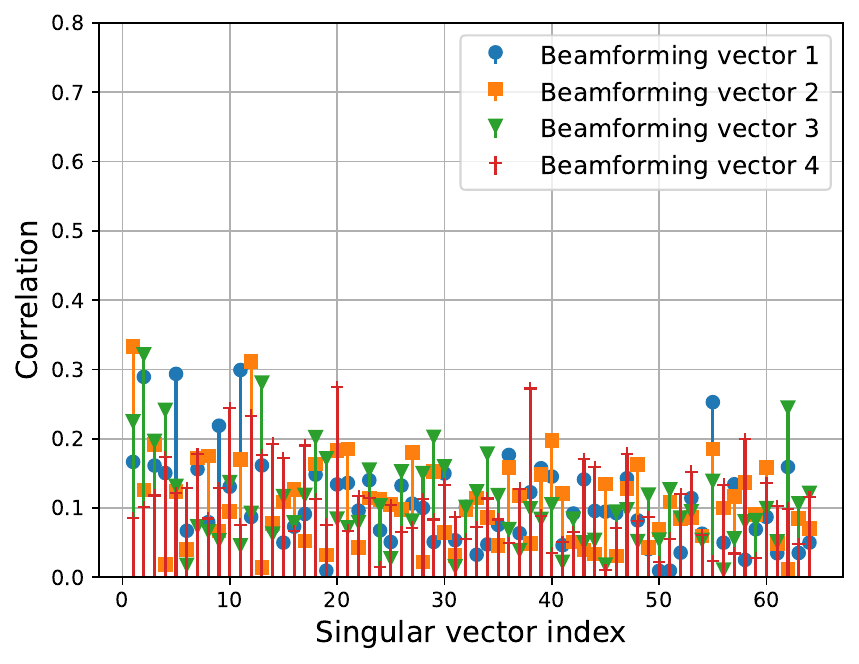}\label{fig:inter_d_d}} 
  \caption{Correlation between the designed data transmission beamformers and the right singular vectors of channel $\bm G$ in a fully digital MIMO system with $M_{\rm t}=M_{\rm r}=64$, $N_{\rm s}=4$ and $L=16$.}
  \label{fig:correlation1}
\end{figure*}
\begin{figure*}
     \centering
     \subfloat[Proposed, SNR=$5$dB, $N_{\rm s}=2$]{\includegraphics[width=0.24\textwidth]{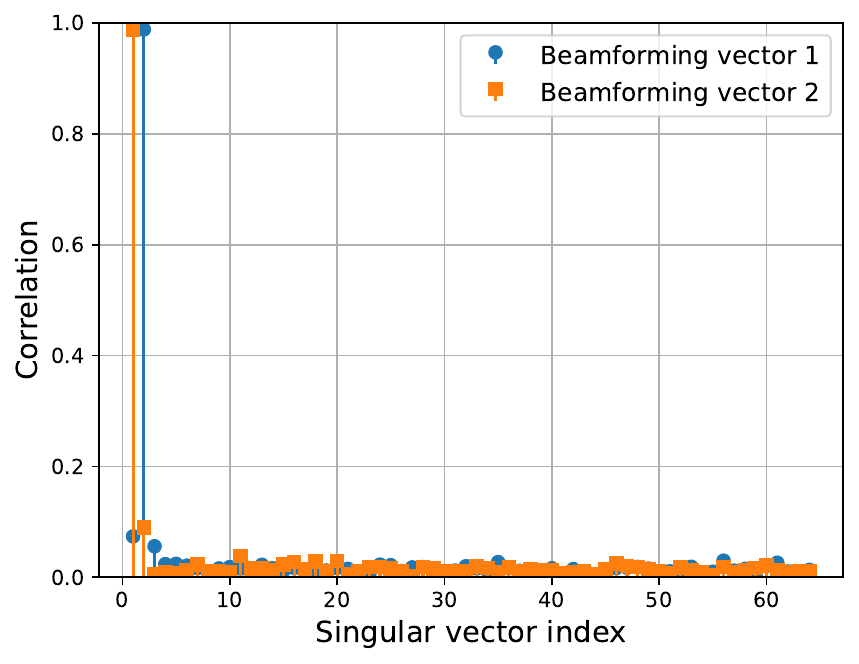}\label{fig:corr_a}}
      \hfil
    \subfloat[Power iteration with codebook, SNR=$5$dB, $N_{\rm s}=2$]{\includegraphics[width=0.24\textwidth]{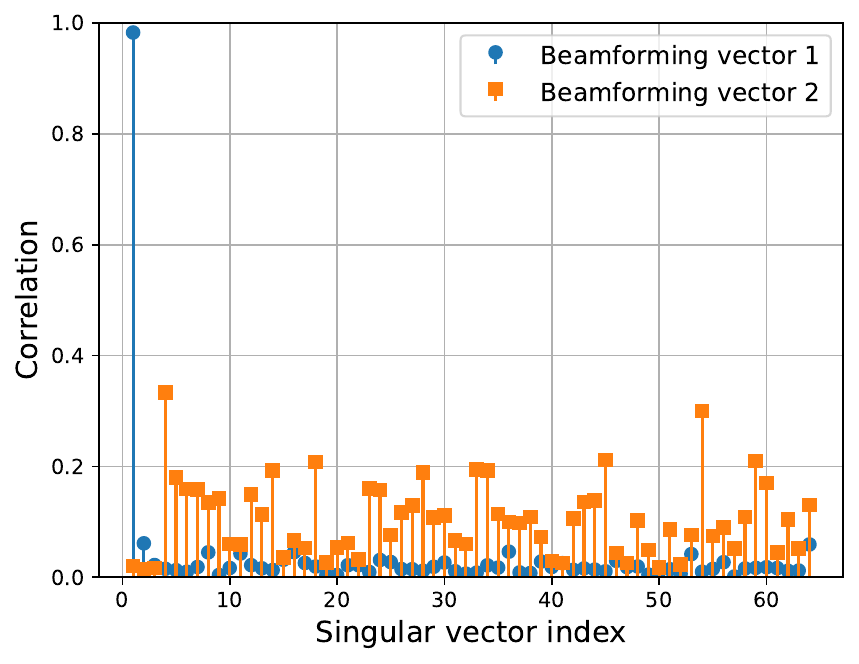}\label{fig:corr_b}}
    \hfil
    \subfloat[Proposed, SNR=$10$dB, $N_{\rm s}=4$]{\includegraphics[width=0.24\textwidth]{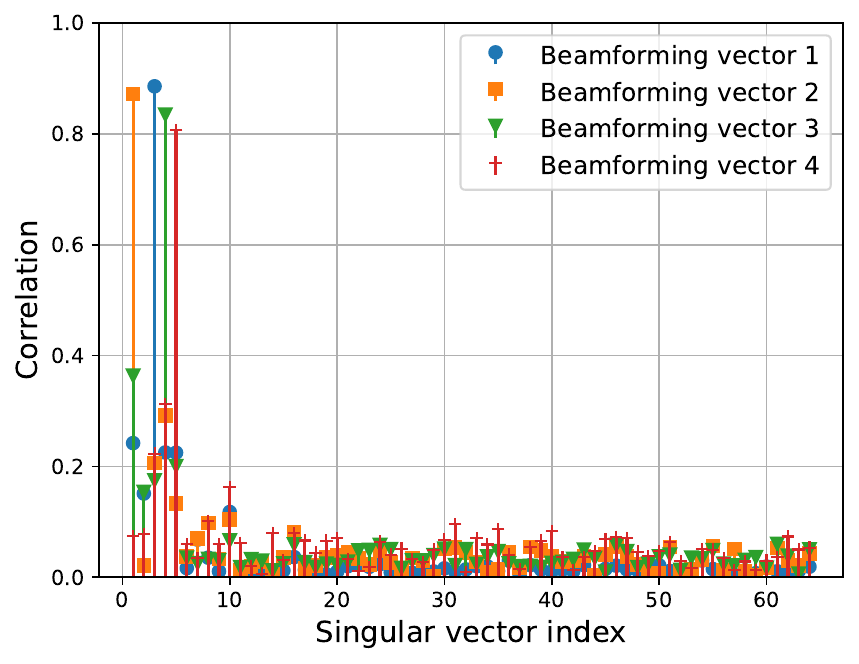}\label{fig:corr_c}}
    \hfil
    \subfloat[Power iteration with codebook, SNR=$10$dB, $N_{\rm s}=4$]{\includegraphics[width=0.24\textwidth]{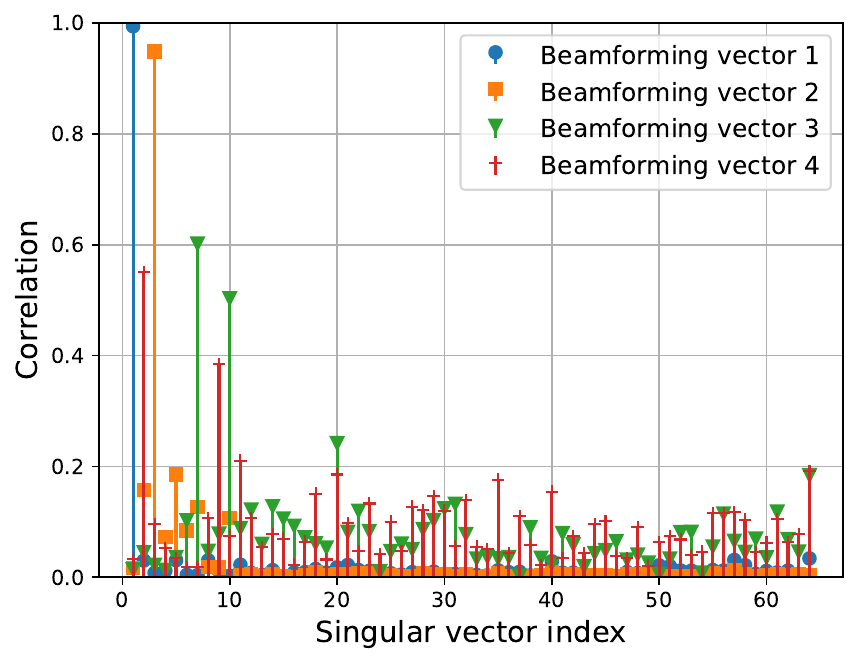}\label{fig:corr_d}}
     \caption{Correlation between the designed data transmission beamformers and the right singular vectors of channel $\bm G$  in a hybrid MIMO system with $M_{\rm t}=M_{\rm r}=64$, $N_{\rm t}^{\rm RF}=N_{\rm t}^{\rm RF}=8$ RF chains and $L=8$.}
     \label{fig:correlation2}
\end{figure*}
So far we have seen the superior performance of the proposed active sensing method both in fully digital MIMO and the hybrid MIMO scenarios. To understand the performance gain better, we compare the proposed active sensing method with the conventional power iteration method in terms of the correlation between the learned solutions and the true singular vectors. 
Specifically, given an estimated singular vector $\bm w$ of the channel, we evaluate its correlation with the true singular vectors as follows:
\begin{equation}
    f(\bm w;\bm u_i)=|\bm w^{\sf H}\bm u_i|, \quad i=1,\ldots,M_{\rm r/\rm t},
\end{equation}
where $\bm u_i$ is the singular vector of $\bm G$ corresponding to the $i$th largest singular value.
We note that similar behavior can be observed among the learned receive beamforming matrix and the transmit beamforming matrix, but we only present the results for the transmit beamformers here.

For the fully digital scenario, we randomly generate a channel realization and plot the correlation between the learned data transmission beamformer and the right singular vectors in Fig.~\ref{fig:correlation1}. There are $4$ beamforming vectors in total, corresponding to the $4$ data streams. By comparing Fig.~\ref{fig:inter_d_a} and Fig.~\ref{fig:inter_d_b}, we observe that the beamforming vectors learned by the proposed active sensing method match better to the top-$4$ singular vectors and have less power leakage to the directions with small singular values. This explains why the proposed methods can achieve better performance. By comparing Fig.~\ref{fig:inter_d_c} and Fig.~\ref{fig:inter_d_d}, which plot the results of the $-5$dB scenario, it can be seen that the proposed method can still output solutions that match the dominant singular vectors but the power iteration method fails to match the dominant singular vector directions.

For the hybrid scenario, we plot the results in Fig.~\ref{fig:correlation2}.  Here, the correlation is calculated between the columns of the overall beamforming matrix $\bm W_{\rm t}$ in \eqref{eq:w_t}, which involves the product of the designed digital and analog beamforming matrices, and the singular vectors. The scenario with $N_{\rm s}=2$ data streams is shown in Fig.~\ref{fig:corr_a} and Fig.~\ref{fig:corr_b}. It can be seen that the beamforming vectors produced by the proposed active sensing method closely match the top-$2$ right singular vectors of the channel, respectively. However, the power method with the codebook only finds the largest singular vector and misses the second-largest one.  The scenario with $N_{\rm s}=4$ data streams is shown in Fig.~\ref{fig:corr_a} and Fig.~\ref{fig:corr_b}, we can observe that the solution obtained by the proposed method can match well with the dominant singular vectors (with a mismatch to the second one), but the solutions produced by the power iteration method with codebook spreads out more power on the other directions. This explains why the performance of the proposed active sensing method is better than the power method with the codebook. 

\subsection{Scalability}
\textcolor{black}{In term of scalability, first note that the learning-based approach is trained
offline, so the main issue is inference complexity. The inference stage primarily 
involves matrix-vector multiplications which can be efficiently implemented in 
hardware such as a graphic processing unit (GPU). The
additional complexity of the proposed active sensing method, as compared to the
power iteration method, arises from processing the $N_{\rm s}$  sets of GRU
cells and DNNs corresponding to the $N_{\rm s}$ data streams. Table
\ref{tab:complexity} shows the run time complexity comparison in a fully
digital massive MIMO system where $M_{\rm t}=128$,  $M_{\rm r}=32$ and $N_{\rm s}=2$. 
It can be observed that the proposed active sensing method achieves
superior performance with comparable run-time as other benchmarks when running on a GPU.}

\begin{table}[t]
  \caption{Run time complexity comparison.}
  \label{tab:complexity}
  \begin{tabular}{|c|c|c|}
    \hline
    Methods &  $\log|\det(\bm W_{\rm r}^{\sf H} \bm G \bm W_{\rm t})|^2$ & CPU inference time (ms)\\ \hline
    LMMSE+SVD& 12.7124 & 0.3901 \\
    Power iteration & 10.7937 & {0.0230 }\\
    Summed power  & 11.1117 & 0.0450 \\
    Proposed  & \textbf{14.9065} & 0.9079 (CPU)/ 0.0348 (GPU) \\ \hline
  \end{tabular}
\end{table}

\textcolor{black}{
We note that there are several ways to further reduce the computational
complexity in practical implementations. First, the $N_{\rm s}$ sets of GRU cells and DNNs can
be executed in parallel, meaning the actual additional complexity is equivalent
to the runtime of a single set of GRU cells and a DNN. Second, we observe that the
proposed active sensing method can achieve satisfactory performance after just
a few ping-pong rounds, allowing for early termination of the algorithm to
conserve computational resources and pilot signals. Finally, while we opt for
relatively large neural networks in the simulations to characterize the ultimate 
performance, we also found that reducing the size of the neural network does
not impact performance significantly.}

\section{Conclusion}\label{sec:conclusion}
This paper considers the problem of precoding and combining matrices design in a TDD massive MIMO system with fully digital or hybrid antenna structures. The conventional design involves a channel estimation stage followed by an optimization step. This can lead to significant pilot training overhead. To address this issue, this paper proposes a learning-based framework to directly design the optimal precoding and combining matrices based on a ping-pong pilot training stage, where the sensing beamformers are actively designed as functions of the pilots received so far. Compared to previous approaches, the proposed algorithm achieves significantly better performance and maintains superior performance in low SNR regimes both in fully digital and hybrid MIMO scenarios. Simulations also show that the proposed algorithm works well in both the multipath channel and Rayleigh fading channel models.

\bibliographystyle{IEEEtran}
\bibliography{refs,IEEEabrv}

\begin{thebibliography}{10}
\providecommand{\url}[1]{#1}
\csname url@samestyle\endcsname
\providecommand{\newblock}{\relax}
\providecommand{\bibinfo}[2]{#2}
\providecommand{\BIBentrySTDinterwordspacing}{\spaceskip=0pt\relax}
\providecommand{\BIBentryALTinterwordstretchfactor}{4}
\providecommand{\BIBentryALTinterwordspacing}{\spaceskip=\fontdimen2\font plus
\BIBentryALTinterwordstretchfactor\fontdimen3\font minus
  \fontdimen4\font\relax}
\providecommand{\BIBforeignlanguage}[2]{{%
\expandafter\ifx\csname l@#1\endcsname\relax
\typeout{** WARNING: IEEEtran.bst: No hyphenation pattern has been}%
\typeout{** loaded for the language `#1'. Using the pattern for}%
\typeout{** the default language instead.}%
\else
\language=\csname l@#1\endcsname
\fi
#2}}
\providecommand{\BIBdecl}{\relax}
\BIBdecl

\bibitem{tao_spawc}
T.~Jiang and W.~Yu, ``Active sensing for reciprocal {MIMO} channels,'' in
  \emph{Proc. IEEE Int. Workshop Signal Process. Advances Wireless Commun.
  (SPAWC)}, Shanghai, China, Sept. 2023.

\bibitem{6375940}
F.~Rusek, D.~Persson, B.~K. Lau, E.~G. Larsson, T.~L. Marzetta, O.~Edfors, and
  F.~Tufvesson, ``Scaling up {MIMO}: Opportunities and challenges with very
  large arrays,'' \emph{{IEEE} Signal Process. Mag.}, vol.~30, no.~1, pp.
  40--60, Jan. 2013.

\bibitem{BJORNSON20193}
E.~Björnson, L.~Sanguinetti, H.~Wymeersch, J.~Hoydis, and T.~L. Marzetta,
  ``Massive {MIMO} is a reality—what is next?: Five promising research
  directions for antenna arrays,'' \emph{Digital Signal Process.}, vol.~94, pp.
  3--20, Nov. 2019.

\bibitem{6515173}
T.~S. Rappaport, S.~Sun, R.~Mayzus, H.~Zhao, Y.~Azar, K.~Wang, G.~N. Wong,
  J.~K. Schulz, M.~Samimi, and F.~Gutierrez, ``Millimeter wave mobile
  communications for {5G} cellular: It will work!'' \emph{IEEE Access}, vol.~1,
  pp. 335--349, May 2013.

\bibitem{7400949}
R.~W. Heath, N.~González-Prelcic, S.~Rangan, W.~Roh, and A.~M. Sayeed, ``An
  overview of signal processing techniques for millimeter wave {MIMO}
  systems,'' \emph{{IEEE} J. Sel. Topics Signal Process.}, vol.~10, no.~3, pp.
  436--453, April 2016.

\bibitem{6717211}
O.~E. Ayach, S.~Rajagopal, S.~Abu-Surra, Z.~Pi, and R.~W. Heath, ``Spatially
  sparse precoding in millimeter wave {MIMO} systems,'' \emph{{IEEE} Trans.
  Wireless Commun.}, vol.~13, no.~3, pp. 1499--1513, March 2014.

\bibitem{1323251}
T.~Dahl, N.~Christophersen, and D.~Gesbert, ``Blind {MIMO} eigenmode
  transmission based on the algebraic power method,'' \emph{{IEEE} Trans.
  Signal Process.}, vol.~52, no.~9, pp. 2424--2431, Sept. 2004.

\bibitem{1193803}
B.~Hassibi and B.~Hochwald, ``How much training is needed in multiple-antenna
  wireless links?'' \emph{{IEEE} Trans. Inf. Theory}, vol.~49, no.~4, pp.
  951--963, April 2003.

\bibitem{9427148}
T.~Jiang, H.~V. Cheng, and W.~Yu, ``Learning to reflect and to beamform for
  intelligent reflecting surface with implicit channel estimation,'' \emph{IEEE
  J. Sel. Areas Commun.}, vol.~39, no.~7, pp. 1931--1945, July 2021.

\bibitem{6847111}
A.~Alkhateeb, O.~El~Ayach, G.~Leus, and R.~W. Heath, ``Channel estimation and
  hybrid precoding for millimeter wave cellular systems,'' \emph{{IEEE} J. Sel.
  Topics Signal Process.}, vol.~8, no.~5, pp. 831--846, Oct. 2014.

\bibitem{8625694}
X.~Song, S.~Haghighatshoar, and G.~Caire, ``Efficient beam alignment for
  millimeter wave single-carrier systems with hybrid {MIMO} transceivers,''
  \emph{{IEEE} Trans. Wireless Commun.}, vol.~18, no.~3, pp. 1518--1533, Mar.
  2019.

\bibitem{8356247}
------, ``A scalable and statistically robust beam alignment technique for
  millimeter-wave systems,'' \emph{{IEEE} Trans. Wireless Commun.}, vol.~17,
  no.~7, pp. 4792--4805, July 2018.

\bibitem{9914567}
W.~Yu, F.~Sohrabi, and T.~Jiang, ``Role of deep learning in wireless
  communications,'' \emph{IEEE BITS Inf. Theory Mag.}, vol.~2, no.~2, pp.
  56--72, Nov. 2022.

\bibitem{8792366}
S.-E. Chiu, N.~Ronquillo, and T.~Javidi, ``Active learning and {CSI}
  acquisition for {mmWave} initial alignment,'' \emph{{IEEE} J. Sel. Areas
  Commun.}, vol.~37, no.~11, pp. 2474--2489, Nov. 2019.

\bibitem{sohrabi2021active}
F.~Sohrabi, T.~Jiang, W.~Cui, and W.~Yu, ``Active sensing for communications by
  learning,'' \emph{{IEEE} J. Sel. Areas Commun.}, vol.~40, no.~6, pp.
  1780--1794, June 2022.

\bibitem{10051966}
T.~Jiang, F.~Sohrabi, and W.~Yu, ``Active sensing for two-sided beam alignment
  using ping-pong pilots,'' in \emph{Asilomar Conf. Signals Syst. Comput.},
  Pacific Grove, California, USA, Nov. 2022, pp. 913--918.

\bibitem{10124207}
------, ``Active sensing for two-sided beam alignment and reflection design
  using ping-pong pilots,'' \emph{{IEEE} J. Sel. Areas Inf. Theory}, vol.~4,
  pp. 24--39, May 2023.

\bibitem{han2023}
H.~Han, T.~Jiang, and W.~Yu, ``Active beam tracking with reconfigurable
  intelligent surface,'' in \emph{Proc. IEEE Int. Conf. Acoust., Speech Signal
  Process. (ICASSP)}, Rhodes, Greece, June 2023.

\bibitem{Zhongze2023}
Z.~Zhang, T.~Jiang, and W.~Yu, ``Localization with reconfigurable intelligent
  surface: An active sensing approach,'' \emph{{IEEE} Trans. Wireless Commun.},
  early access, 2023.

\bibitem{7947217}
D.~Ogbe, D.~J. Love, and V.~Raghavan, ``Noisy beam alignment techniques for
  reciprocal {MIMO} channels,'' \emph{IEEE Trans. Signal Process.}, vol.~65,
  no.~19, pp. 5092--5107, Oct. 2017.

\bibitem{4203059}
T.~Dahl, S.~Silva~Pereira, N.~Christophersen, and D.~Gesbert, ``Intrinsic
  subspace convergence in {TDD MIMO} communication,'' \emph{{IEEE} Trans.
  Signal Process.}, vol.~55, no.~6, pp. 2676--2687, June 2007.

\bibitem{7575663}
E.~de~Carvalho and J.~B. Andersen, ``Transceivers based on ping-pong beam
  training for large multiuser {MIMO} systems,'' \emph{{IEEE} Wireless Commun.
  Lett.}, vol.~5, no.~6, pp. 656--659, 2016.

\bibitem{7556971}
P.~Xia, R.~W. Heath, and N.~Gonzalez-Prelcic, ``Robust analog precoding designs
  for millimeter wave {MIMO} transceivers with frequency and time division
  duplexing,'' \emph{{IEEE} Trans. Commun.}, vol.~64, no.~11, pp. 4622--4634,
  Nov. 2016.

\bibitem{7439748}
H.~Ghauch, T.~Kim, M.~Bengtsson, and M.~Skoglund, ``Subspace estimation and
  decomposition for large millimeter-wave {MIMO} systems,'' \emph{{IEEE} J.
  Sel. Topics Signal Process.}, vol.~10, no.~3, pp. 528--542, April 2016.

\bibitem{7996580}
C.~N. Manchón, E.~de~Carvalho, and J.~B. Andersen, ``Ping-pong beam training
  with hybrid digital-analog antenna arrays,'' in \emph{Proc. IEEE Int. Conf.
  Commun. (ICC)}, Paris, France, May 2017.

\bibitem{8644444}
N.~Akdim, C.~N. Manchón, M.~Benjillali, and E.~de~Carvalho, ``Ping pong beam
  training for multi stream {MIMO} communications with hybrid antenna arrays,''
  in \emph{Proc. IEEE Global Commun. Conf. (Globecom) Workshops}, Abu Dhabi,
  UAE, Dec. 2018.

\bibitem{cover_elements_2006}
T.~M. Cover and J.~A. Thomas, \emph{Elements of Information Theory},
  2nd~ed.\hskip 1em plus 0.5em minus 0.4em\relax Hoboken, NJ:
  Wiley-Interscience, 2006.

\bibitem{cho2014properties}
K.~Cho, B.~van Merrienboer, {\c{C}}.~G{\"{u}}l{\c{c}}ehre, D.~Bahdanau,
  F.~Bougares, H.~Schwenk, and Y.~Bengio, ``Learning phrase representations
  using {RNN} encoder-decoder for statistical machine translation,'' in
  \emph{Proc. Conf. Empirical Methods Natural Lang. Process. (EMNLP)}, Doha,
  Qatar, Oct. 2014, pp. 1724--1734.

\bibitem{he2016deep}
K.~He, X.~Zhang, S.~Ren, and J.~Sun, ``Deep residual learning for image
  recognition,'' in \emph{Proc. IEEE Conf. on Comput. Vision and Pattern
  Recognit. (CVPR)}, Las Vegas, Nevada, USA, June 2016, pp. 770--778.

\bibitem{Paszke2019PyTorch}
A.~{Paszke}, S.~{Gross}, S.~{Chintala}, G.~{Chanan}, E.~{Yang}, Z.~{DeVito},
  Z.~{Lin}, A.~{Desmaison}, L.~{Antiga}, and A.~{Lerer}, ``Pytorch: An
  imperative style, high-performance deep learning library,'' in \emph{Proc.
  Conf. Neural Info. Process. Systems (NeurIPS)}, Vancouver, Canada, Dec. 2019,
  pp. 8024--8035.

\bibitem{kingma2015adam}
D.~P. {Kingma} and J.~{Ba}, ``{Adam: A Method for Stochastic Optimization},''
  in \emph{Int. Conf. Learning Representations (ICLR)}, Y.~{Bengio} and
  Y.~{LeCun}, Eds., San Diego, CA, USA, May 2015, pp. 1--13.

\bibitem{sionna}
J.~Hoydis, S.~Cammerer, F.~{Ait Aoudia}, A.~Vem, N.~Binder, G.~Marcus, and
  A.~Keller, ``Sionna: An open-source library for next-generation physical
  layer research,'' \emph{arXiv preprint}, Mar. 2022.

\bibitem{7397861}
X.~Yu, J.-C. Shen, J.~Zhang, and K.~B. Letaief, ``Alternating minimization
  algorithms for hybrid precoding in millimeter wave {MIMO} systems,''
  \emph{{IEEE} J. Sel. Topics Signal Process.}, vol.~10, no.~3, pp. 485--500,
  April 2016.

\bibitem{7389996}
F.~Sohrabi and W.~Yu, ``Hybrid digital and analog beamforming design for
  large-scale antenna arrays,'' \emph{{IEEE} J. Sel. Topics Signal Process.},
  vol.~10, no.~3, pp. 501--513, Apr. 2016.

\bibitem{4385788}
J.~A. Tropp and A.~C. Gilbert, ``Signal recovery from random measurements via
  orthogonal matching pursuit,'' \emph{{IEEE} Trans. Inf. Theory}, vol.~53,
  no.~12, pp. 4655--4666, Dec. 2007.

\end{thebibliography}

\end{document}